\documentclass[runningheads]{llncs}
\usepackage[T1]{fontenc}
\usepackage[utf8]{inputenc}
\usepackage{lmodern}
\usepackage{microtype}
\usepackage{graphicx}
\usepackage{amssymb}
\usepackage{xcolor}
\usepackage{amsmath}
\usepackage{booktabs}
\usepackage{tabularx}
\usepackage{xltabular}
\usepackage{array}
\newcolumntype{N}{>{\hsize=0.9\hsize\raggedright\arraybackslash}X}
\newcolumntype{U}{>{\hsize=0.7\hsize\raggedright\arraybackslash}X}
\newcolumntype{D}{>{\hsize=1.4\hsize\raggedright\arraybackslash}X}
\usepackage{natbib}
\setcitestyle{authoryear,round,aysep={}}

\usepackage[hidelinks]{hyperref}

\begin{document}
\title{Understanding CEO-Employee Compensation Dynamics: An Agent-Based Simulation Approach}
\titlerunning{Understanding CEO-Employee Compensation Dynamics}
\author{Darío Blanco-Fernández\orcidID{0000-0002-8275-701X}}
\authorrunning{D. Blanco-Fernández}
\institute{Department of Economics, University of Oviedo, Spain\\
\email{blancodario@uniovi.es}}
\maketitle
\begin{abstract}
CEO compensation has often been examined from the top down, i.e., as the outcome of a negotiation between the CEO, the board of directors, and the shareholders. While these approaches have provided valuable insights into the nature of CEO compensation and its relations with firm performance and intra-firm inequality, we believe they leave aside the internal decision-making dynamics of CEOs and employees. To fill this research gap, we develop an agent-based model built on the NK framework in which a CEO compensated for long-term firm performance interacts with several employees rewarded for short-term departmental performance. This temporal dichotomy in their respective compensation functions creates a potential temporal misalignment of incentives. By varying the horizon considered by the CEO to make their decisions, we examine how this temporal (mis)alignment of incentives shapes firm performance and pay inequality between the CEO and the employees. Our findings suggest that aligning the CEO's horizon with the employees' short-term incentives yields the highest performance, raising both the level attained and the speed of improvement across all task environments examined. This finding contrasts with the usual advice given in CEO compensation research, which is to avoid short-termism. Furthermore, the model shows that the longest horizon examined produces both the lowest performance and the highest pay inequality. This pattern, commonly attributed to managerial power, emerges in our model from incentive misalignment alone, without agents holding any power over their own compensation.

\end{abstract}
\section{Introduction}

Executive compensation is both a governance device, meant to align managerial decisions with the maximisation of long-term firm value, and one of the most visible sources of pay inequality within firms. That inequality has grown markedly. Among S\&P 500 firms, the ratio of CEO pay to average worker pay rose from about 40 in 1980 to 335 in 2015 \citep{Edmans2017}. Yet, prior research has not produced a consensus on why CEO compensation has grown so much faster than the pay of other employees or how it relates to firm performance \citep{Edmans2016}. These questions are highly relevant due to their connections to wider debates on income inequality and its policy implications \citep{Edmans2016, Keller2021,Piketty2014}.

The literature on CEO compensation is broadly divided into two perspectives. The first, the \textit{shareholder value view}, sees CEO compensation growth as a consequence of market forces and a product of optimal contracting \citep{Edmans2016}. According to this view, shareholders and directors design compensation schemes that incentivise the CEO to increase long-term firm value \citep{Edmans2017,Edmans2023}. Thus, the increase in CEO compensation would reflect an overall improvement in performance indicators such as financial results \citep{Albuquerque2024,Bradley2013} and firm size \citep{Edmans2016,Frydman2018}. 

Conversely, the \textit{rent extraction view} disputes the interrelations between incentive alignment, CEO compensation, and firm performance \citep{Edmans2017}. Instead, researchers taking this perspective argue that CEO compensation is influenced by the top executives themselves, who exploit weak corporate governance \citep{Boyd1994,Edmans2023}, opaque corporate networks \citep{Henriksen2025}, and favourable economic conditions \citep{Keller2021}. Therefore, this view argues that expanded managerial power, rather than firm performance, is the key driver behind CEO compensation growth \citep{Edmans2017}.

Although their interpretations of CEO compensation dynamics differ, both perspectives can be described as \textit{top-down} approaches. Specifically, both describe executive compensation as a product of a negotiation process between the CEO, the shareholders, and the board of directors \citep{Edmans2017}. While both have proven valuable in examining some of the key factors behind the growth in CEO compensation, they often leave the incentives of those below the CEO outside their scope of analysis. Yet firm performance results from decisions taken at all levels of the firm, with these decisions being shaped by the decision-makers' pay structure \citep{Bandiera2007,Lazear2000,Siggelkow2005}. Since CEOs and lower-level employees are typically paid under very different schemes \citep{Edmans2017,Mueller2017}, their incentives do not necessarily point in the same direction. Consequently, to explain the growing concentration of pay at the top of the firm, we need to consider how the CEO's incentives interact with those of the lower-level decision-makers, and how both performance and the distribution of pay emerge from these interactions, i.e., from the \textit{bottom up}.

To account for these interrelations, we propose an Agent-Based Model (ABM) based on the \textit{NK} framework and adapted to model hierarchical decision-making \citep{Siggelkow2005}. Our model captures the interactions between a CEO and lower-level employees, building on the potential incentive misalignment that may appear during their decision-making processes. This incentive misalignment emerges from the fundamental differences in pay structure between CEOs and lower-ranked employees: the former are usually compensated using variable packages based on long-term firm performance maximisation, while the latter mostly receive fixed salaries combined with smaller, short-term performance-based commissions or bonuses \citep{Edmans2017,Lazear2000}. We explicitly model this incentive misalignment by endowing the CEO with the ability to focus on short-, medium-, or long-term compensation maximisation, as opposed to the employees---who only focus on the short term. This temporal conceptualisation extends prior simulation-based research on incentive misalignment, which has mainly focused on the misalignment between departmental and firm-wide incentives \citep{Rivkin2002,Siggelkow2005,Wall2016}. This paper therefore contributes both to the literature on CEO compensation, by showing how the pay gap between the CEO and lower-level employees emerges from the interaction of their decisions, and to simulation-based organisational research, by adding a temporal dimension to the study of incentive misalignment within hierarchies.

The temporal misalignment of incentives motivates the paper's two research questions. First, how does the temporal misalignment between the CEO's and the employees' incentives influence the relationship between CEO compensation and firm performance? Second, how do these interrelations shape pay inequality between the CEO and the employees? Our results suggest that aligning the CEO's decision-making horizon with the employees' short-term incentives improves performance, which provides a contrast to the conventional wisdom often found in the literature \citep{Edmans2017}. Moreover, the longest horizon examined produces both the lowest performance and the highest pay inequality observed. This combination, usually taken as a sign of managerial power \citep{Edmans2017,Henriksen2025}, emerges in our model only from the misalignment between the CEO's and the employees' incentives.

\section{Background}
\subsection{Stylised facts of CEO and employee compensation}
\leavevmode\label{sec:stylised}The specifics of CEO pay may differ between firms, but most of them follow certain common principles. Compensation packages often consist of a fixed wage plus a set of \textbf{variable components}, including bonuses, long-term incentive-based payments, options, and stocks \citep{Edmans2017}. These variable components are normally performance-dependent, i.e., they are tied to certain performance measures and are not triggered until the firm reaches a particular threshold \citep{Albuquerque2024,Edmans2017,Frydman2010}. Moreover, certain bonuses are approved at the discretion of the board of directors \citep{DeAngelis2020}. 

The relative importance of the variable components has increased over time \citep{Albuquerque2024}. In 1992, salaries represented approximately 42\% of the total earnings of the average S\&P 500 CEO \citep{Edmans2017}. By 2014, variable components represented more than 80\% of these total earnings. Generally, these variable components include stocks (44\% of earnings), options (16\%), and bonuses (23\%), including long-term incentive plans. By contrast, the salary only represented 13\% of what the average S\&P 500 CEO earned in 2014. This shift to variable, performance-based pay has effectively transformed the role of the CEO from merely the highest-level employee of the firm to a mixed figure that combines elements of a top employee and a shareholder. 

This transformation is far from arbitrary, and builds on the idea of \textit{incentive alignment} between the firm's CEO and the shareholders \citep{Chng2012,Frydman2010,Jaggia2022}. According to this logic, a CEO who has a pecuniary interest in a firm is expected to make decisions that improve the long-term value of the firm, which is what shareholders are assumed to want \citep{Edmans2017,Frydman2010}. Hence, CEO compensation packages are usually \textbf{dependent on long-term performance} measures \citep{Albuquerque2024,Edmans2017}. 

Nevertheless, the relationship between firm performance and CEO compensation is asymmetric \citep{Edmans2023}. Pay increases with performance, even when improvements in performance reflect favourable external conditions rather than the CEO's actions \citep{Frydman2010,Keller2021}. By contrast, comparable declines are not followed by reductions in pay. Directors themselves report that good recent performance justifies raising compensation, while poor performance does not justify reducing it \citep{Edmans2023}. Consequently, CEO compensation is \textbf{downwardly rigid}, i.e., it is able to capture any improvements in performance, irrespective of its source, while avoiding the negative consequences of poor performance \citep{Taylor2013}.

The pay structure of lower-level employees follows other principles. Generally, the further down we go in a firm's hierarchy, \textbf{the less variable is compensation} \citep{Edmans2017}. Median employee compensation varies little with firm performance, which evidences that employees located down in the firm's hierarchy earn mostly fixed wages. This reliance on fixed wages also implies that employee compensation is downwardly rigid, although this rigidity may be linked to the existence of legal and institutional agreements regulating salaries \citep{Henriksen2025}. Furthermore, if variable components exist, they are usually \textbf{dependent on short-term performance} \citep{Bandiera2007,Lazear2000}, rather than on long-term firm value. 

These structural differences in how CEOs and lower-level employees are compensated offer one explanation for the growth of pay inequality within firms. 
Pay structure matters because these two kinds of compensation respond differently to the same outcome. While lower-level employee compensation responds weakly to firm performance, CEO compensation is highly sensitive to increases in performance \citep{Edmans2017,Taylor2013}. Thus, in the face of increasing value, size, or productivity, the pay gap between CEOs and lower-level employees should widen \citep{Edmans2017,Mueller2017}. Higher within-firm inequality is often associated with larger and better-performing firms \citep{Mueller2017}, yet increases in CEO pay are also followed by lower wage growth for the firm's workers \citep{Henriksen2025}. Whether this divergence reflects the efficient pricing of managerial talent---as the shareholder value view argues---or the CEO's ability to capture value at the expense of employees or shareholders---as the rent extraction view argues---remains contested. 

\subsection{Incentive alignment beyond the CEO–shareholder relationship}
\leavevmode\label{sec:incentive}Both the shareholder value and the rent extraction views explain the divergence in executive and employee compensation through the interrelations between the CEO, the board of directors, and the shareholders \citep{Edmans2017}. Hence, the principal-agent model \citep{Hlmstrom1979} has served as the main theoretical framework for examining how CEO compensation is set and its evolution over time \citep{Albuquerque2024}. In its standard formulation, a principal delegates a task to an agent whose actions it cannot observe, tying the agent's compensation to an observable outcome \citep{Hlmstrom1979}. In an efficient contract, this outcome is any measure that is informative about the agent's actions, as they will exert more or less effort depending on their aversion to risk and expected utility \citep{Hlmstrom1979}.

Within our context, the shareholders act as the principal, who offer---through the board of directors---a specific compensation scheme to the CEO, i.e., the agent \citep{Edmans2017,Frydman2010}. An efficient contract should ensure that the CEO takes actions that serve the shareholders' objectives, i.e., the maximisation of long-term firm value \citep{Edmans2012,Edmans2017}. Optimal contracting also involves compensating the CEO for bearing pay risk if risk aversion exists, so it should balance incentive alignment with risk-sharing \citep{Edmans2017,Hlmstrom1979}. Furthermore, an optimal contract should prevent tying pay to external factors that may increase or decrease firm performance \citep{Frydman2010}. Hence, outcomes should be evaluated relatively to the performance of other firms, in order to filter out political or economic shocks outside the scope of action of the CEO \citep{Edmans2017}. 

Because this contracting involves observing outcomes for an extended period of time, an optimal contract must also consider the temporal dynamics inherent to shareholder-CEO incentive alignment. The challenge, therefore, involves designing a compensation scheme that prevents CEOs from engaging in short-termism---i.e., making decisions that inflate short-term performance at the expense of long-term firm value \citep{Edmans2012,Edmans2016}---or in ratcheting---i.e., withholding short-term performance to avoid higher performance targets in future periods \citep{Bouwens2011}. \cite{Edmans2012} argue that rewards should be spread over time, with pay becoming more sensitive to performance as retirement---or end of contract---approaches. Furthermore, \cite{Edmans2016} argue that compensation should be contingent on performance after retirement---or end of contract---in order to prevent short-termism. 

However, the empirical evidence on these insights remains ambiguous \citep{Frydman2010}. \cite{Edmans2017} argue that the theoretically optimal contract is typically highly nonlinear and never observed in practice. Hence, the principal–agent framework serves better as a theoretical benchmark than as a description of practice, as actual contracts converge with it in some respects and depart from it in others. For instance, the risk premium---a central element of the principal-agent model---is often economically small \citep{Albuquerque2024}, and both shareholders and directors rank the CEO's risk attitude and the firm's riskiness among the least important determinants of incentives \citep{Edmans2023}. Additionally, evidence on the use of relative performance evaluation is mixed \citep{Edmans2017}. When benchmarked against a custom peer group, these rewards appear to filter out common shocks effectively, as theory predicts \citep{Bizjak2022}. However, the use of relative performance evaluation is still limited, a fact that proponents of the rent extraction view have employed as evidence for their arguments \citep{Edmans2023}.

Evidence is also mixed on making compensation dependent on long-term performance. \cite{Edmans2023} argue that actual CEO compensation is based on relatively short performance horizons. Moreover, there appears to be disagreement on the effects of increasing these horizons, with shareholders arguing that they could improve firm performance and directors believing that they are long enough \citep{Edmans2023}. Regarding this debate, prior evidence suggests that long-term incentives may indeed improve long-term performance at the cost of short-term outcomes \citep{Edmans2017}.

Beyond these empirical discrepancies, the standard principal–agent framework applied to CEO compensation research also fails to capture several relevant intra-firm dynamics. For instance, the CEO's actions may directly affect firm value, but can also affect it indirectly by changing the optimal actions of their subordinates \citep{Edmans2016}. As the principal-agent framework involves modelling a single agent at two levels, these complementarities between the CEO and other employees remain outside its scope \citep{Edmans2016}. \cite{Edmans2017} argue that placing the CEO decision-making process in a multiagent team setting could be used to address this gap, leading to further insights on internal pay inequality. Moreover, most models assume fully rational agents, leaving aside behavioural aspects such as bounded rationality \citep{Edmans2017}. Overall, these highlighted research gaps raise an issue the standard framework cannot address: how differences in the temporal incentives of CEOs and lower-level employees shape firm performance and pay inequality.

\subsection{Computational approaches to organisational and incentive design}
\leavevmode\label{sec:computational}Simulation-based approaches can be used to relax some of the most restrictive assumptions of the principal-agent model, allowing us to cover some of the research gaps discussed in the previous subsection \citep{Wall2021}. For instance, \cite{Reinwald2024} develop an ABM adaptation of the principal-agent framework that allows for the inclusion of bounded rationality, limited oversight, and restricted information. However, the principal-agent framework remains dyadic even with these inclusions. To answer our research questions, we need a framework that allows us to explicitly model the interactions between the CEO and several lower-level employees and examine their consequences. The NK model for organisational decision-making provides such a framework.

In its original formulation for evolutionary biology, the NK model was employed to examine how the interactions among the individual genes of a species determined its adaptive search process, and the population-level dynamics that emerged from it \citep{Kauffman1993}. Later on, \cite{Levinthal1997} adapted this framework to explain diversity in organisational design as an emergent outcome of the evolution of interdependent organisational attributes. Subsequent work has used this framework to answer questions related to organisational decision-making \citep{Rivkin2002,Wall2023}, the emergence of task formation \citep{Wall2018}, and coordination within self-organised teams \citep{Blanco-Fernndez2025}. Two further lines of research tie directly to our study.

First, the NK model has been used to examine how firm-level outcomes emerge from the interaction of decision-makers within the firm. \cite{Siggelkow2005} compare a range of organisational archetypes that differ in the decision-making power granted to department heads, and show how these design choices shape firm performance. We build on their hierarchical archetype, in which department heads propose alternatives for their departmental policies and the CEO combines these proposals into a firm-wide strategic policy (see Sec. \ref{sec:model}). Furthermore, we extend their approach by conceptualising the CEO not as a neutral coordinator who maximises firm performance, but as an agent of the model who maximises their own compensation. 

Second, researchers have often employed the NK framework to address the issue of within-firm incentive alignment \citep{Wall2021}. These studies typically compare the performance outcomes of departmental and firm-wide incentives. Generally, firm-wide incentives are shown to speed up performance improvement compared to departmental incentives, as they discourage decision-makers from blocking alternatives which improve firm performance at their own expense \citep{Rivkin2002,Siggelkow2005}. However, this positive effect is contingent on the interdependencies between departments \citep{Wall2016}, and departmental incentives may be used to prevent firms from stagnating at suboptimal policies \citep{Dosi2003}. Although these studies shed light on incentive alignment, few have examined how differences in the temporal dimension of the incentives, i.e., in the period over which performance is rewarded, shape firm outcomes.
\section{Model description}
\leavevmode\label{sec:model}Following the ODD protocol and its guidelines \citep{Grimm2020}, we provide a detailed model description in the Appendix. This section contains a summary of the model, its main characteristics and processes. The \textit{purpose} of this model is to examine how the temporal (mis)alignment between the CEO's and employees' incentive schemes shapes firm performance and the divergence in compensation growth. Our approach conceptualises these two latter aspects as emergent outcomes---i.e., from the bottom up---, which contrasts with the top-down, principal-agent framing usually taken to examine executive compensation \citep{Edmans2017,Hlmstrom1979}. In this model, we build a hierarchical firm that searches an \textit{NK} performance landscape. The firm is formed by two kinds of boundedly rational agents: the CEO and the employees. While CEO compensation is based on long-term firm performance maximisation, employees are paid based on their short-term contributions to firm performance. 

\leavevmode\label{sec:patterns}We contrast the model against two \textit{patterns} usually found in the literature. First, that incentive alignment is associated with faster performance improvement \citep{Rivkin2002,Siggelkow2005,Wall2016}. Second, that CEO compensation grows faster than that of other employees, including other senior executives \citep{Edmans2016,Henriksen2025,Mueller2017,Piketty2014}.

The model includes two individual \textit{entities}. The \textbf{CEO} is a single agent who is located at the highest level of a hierarchy. We also model a population of $A=30$ \textbf{employees} that are randomly assigned to one of the $M=3$ \textbf{departments}. The employees compete with their department colleagues to become \textbf{head of department} (see Submodel \ref{sec:hod_selection}). Together, the CEO, the heads of department, and the remaining employees form a hierarchical \textbf{firm} (see Sec. \ref{sec:collectives}). This firm faces a particular \textbf{task environment} based on the \textit{NK} performance landscape \citep{Levinthal1997}, as outlined in Submodel \ref{sec:nkmodel}. This task environment is akin to a \textbf{firm's strategic policy}, which can be partitioned into $M=3$ \textbf{departmental policies}. 

The \textit{state variables} are summarised in Table \ref{tab:app_state_variables} in the Appendix. The CEO is characterised by a compensation received at the end of each period $u_t(\mathrm{CEO})$, the compensation they can project over $H$ periods $\hat{U}_t(\mathbf{d}_t)$, and the performance threshold they face $\bar{C}^{*}$. Employees are assigned to one department $m$, can take a role of head of department, receive compensation $u_t(m)$, have a set of known departmental policies $\Omega_{mt}$, and beliefs about these departmental policies $p^{(i)}_{mt}$ and $q^{(i)}_{mt}$. The NK task environment results in a strategic policy $\mathbf{d}_t$ formed by $N=12$ individual decisions $d_i$ and evolves over time $T$. Each decision contributes $c_i$ to firm performance $C_t(\mathbf{d}_t)$. As for the \textit{scales} of the model, there is no spatial dimension, and time is modelled as a series of discrete periods $t$---representing strategic reporting cycles, such as a quarter---that end at $T=100$.

\begin{figure}[!htbp]
    \centering
    \includegraphics[width=1\textwidth]{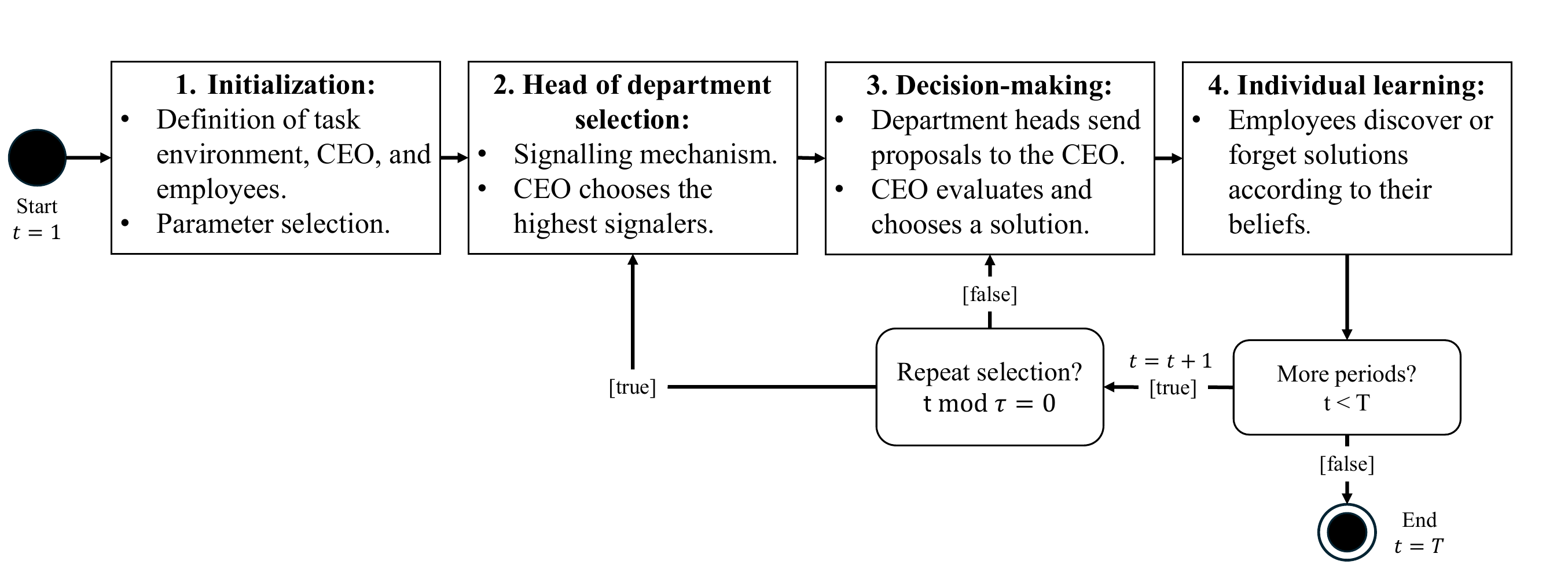}
    \caption{Sequence of events of the model}
    \label{fig:events}
\end{figure}

We structure the model around four \textit{processes}, which correspond to the sequence of events (see Fig. \ref{fig:events}). 

\begin{itemize}
    \item First, during the \textbf{model initialisation}, we determine which scenarios to examine, randomly assign each employee to one of the $M=3$ departments, and generate a baseline performance threshold $\bar{C}^{*}$. Then the main sequence of events starts, which is repeated recursively until it reaches $t=100$. 
    \item The second process is the \textbf{head of department selection}. This process consists of two actions. The first action involves the employees, who estimate the compensation they will receive for each departmental policy known and signal their highest estimated compensation to the CEO. Then, the CEO names as head of department the signaller with the highest estimated compensation for each of the $M=3$ departments. This process is not repeated at every period. Instead, each head of department takes this role during the incoming $\tau$ periods. Once they reach this term limit, the selection process is repeated again, and a new head of department (or the same) is chosen for the following $\tau$ periods.
    \item The third process, \textbf{decision-making}, occurs immediately afterwards, and also involves two actions. The heads of department evaluate all solutions known according to their estimated compensation, and send the $PROP=2$ highest-compensating proposals to the CEO. The CEO then forms $COMP=2$ candidate policies by drawing one of the $PROP=2$ proposals per department at random. Then, the CEO projects these $COMP=2$ candidate policies over $H$ periods, and chooses the policy with the highest projected compensation. However, this policy is only finally implemented if it improves compensation compared to the currently implemented policy over the following $H$ periods. Otherwise, the CEO keeps the status quo, i.e., $\mathbf{d}_t=\mathbf{d}_{t-1}$. Firm performance is then calculated and both the CEO and the employees receive their corresponding compensation. 
    \item Finally, the fourth process is the employees' \textbf{individual learning}. During this process, the employees go through three subprocesses. First, the heads of department alter their beliefs about the solutions known according to their received compensation. Second, the knowledge of all employees wears out as time goes by. Third, guided by their beliefs, all employees may discover a new departmental policy, or forget one that was stored in their memory.
\end{itemize}

The model includes several \textit{design concepts} worth discussing. The model is built on the \textit{basic principle} of the \textbf{\textit{NK} performance landscape} \citep{Levinthal1997}, in which a firm faces a particular complex task---in this case, finding and implementing an optimal firm strategic policy. The dynamics of the model follow the principle of \textbf{incremental search}, by which the firm searches step-by-step for an optimal solution. Moreover, the firm's internal architecture rests on the idea of the \textbf{hierarchical organisational archetype} \citep{Siggelkow2005}, as it is formed by baseline employees, heads of department, and a CEO who is the final decision-making authority. The different agents have \textbf{misaligned incentives} derived from differences in their \textbf{compensation packages}.

These basic principles also influence some of the most relevant \textit{design concepts} of the model. First, both firm performance and the growth in CEO-to-employee relative compensation are \textit{emergent} properties of these principles. Second, employees, the heads of department, and the CEO all have direct objective-seeking \textit{adaptive behaviours}. Their \textbf{objective} is to maximise their compensation, although they differ on their associated compensation function. Employees earn a fixed wage $w_{emp}$, while heads of department earn this wage plus a short-term performance based bonus $u_t(m) = w_{emp}+\frac{1}{S} \sum_{n \in \mathbf{d}_m} c_n(\mathbf{d}_t)$.\footnote{For the sake of simplicity, we set all fixed wages equal to zero, i.e., $w_{CEO}=w_{emp}=0$.} By contrast, the CEO earns a fixed wage plus a variable component based on long-term performance, following:

\begin{equation}\label{eq:ceo_utility}
    u_t(CEO) = \begin{cases}
        w_{CEO} \textrm{ if } C(\mathbf{d}_t) \leq \theta\bar{C}^* \\
        w_{CEO} + \frac{C(\mathbf{d}_t)}{T - t+1} \textrm{ if } C(\mathbf{d}_t) > \theta\bar{C}^* 
    \end{cases}
\end{equation}

Because the variable component weighs performance in later periods more heavily, the CEO is incentivised to maximise their long-term compensation, rather than only immediate gains. Hence, we endow the CEO with the capacity to (imperfectly) estimate their compensation over $H$ periods (see Submodel \ref{sec:decision-making}). Consequently, the CEO's \textbf{objective} is to maximise compensation over an $H$-period horizon, and the objective function becomes:

\begin{equation}\label{eq:ceo_objective}
    \hat{U}_t(\mathbf{d}_t) = \sum_{h=0}^{H-1}
    \begin{cases}
        w_{CEO}  \text{ if } C(\mathbf{d}_{t+h}) \leq \theta\bar{C}^{*}, \\[4pt]
        w_{CEO} + \dfrac{C(\mathbf{d}_{t+h})}{T-(t+h)+1} & \text{if } C(\mathbf{d}_{t+h}) > \theta\bar{C}^{*}.
    \end{cases}
\end{equation}

Finally, employees \textit{learn} over time by discovering and forgetting solutions following a reinforcement learning mechanism with Bayesian updates of beliefs \citep{Blanco-Fernndez2025}.

\subsection{Experimental design and parameters}
Our four main experimental parameters are: the interdependencies $K$, the interdependence structure (see Fig. \ref{fig:matrix} in the appendix), the head of department tenure $\tau$, and the CEO forecasting horizon $H$. Together, these four parameters form 36 scenarios that we examine individually. Each scenario is replicated over $R=1500$ runs.

\begin{table}[!htbp]
\centering
\caption{Parameters varied across scenarios.}
\label{tab:scenarios}
\begin{tabularx}{\textwidth}{@{}>{\raggedright\arraybackslash}p{2.9cm} l >{\raggedright\arraybackslash}p{1.9cm} >{\raggedright\arraybackslash}X@{}}
\toprule
\textbf{Parameter} & \textbf{Symbol} & \textbf{Values} & \textbf{Interpretation} \\
\midrule
Interdependencies & $K$ & $3;\ 5$
  & Complexity of the task environment. \\[2pt]
Interdependence structure & --- & block-diagonal; random
  & Whether interdependencies fall within departments or also across them. \\[2pt]
CEO forecasting horizon & $H$ & $1;\ 10;\ 100$
  & Number of future periods over which the CEO evaluates a candidate policy. \\[2pt]
Turnover rate & $\tau$ & $1;\ 10;\ \emptyset$
  & Number of periods between successive head of department selections. Under $\tau=\emptyset$ heads are selected once, at $t=1$, and retain the role for the entire run. \\
\bottomrule
\end{tabularx}
\end{table}

The central experimental variable in this model is $H$, the CEO's forecasting horizon. Under $H=1$, the CEO forecasts compensation only in the immediate period, which we refer to as \textit{short-term forecasting}. This also means that there is perfect temporal alignment between the CEO and the heads of department. We denote $H=10$ as \textit{medium-term forecasting} and $H=100$---which effectively means that the CEO forecasts the entire simulation temporal horizon $T=100$---as \textit{long-term forecasting}. In these two scenarios, the CEO forecasts with imperfect information, as they have to anticipate the employees' future proposals and learning paths. Hence, the higher $H$, the higher the risk of making errors in the forecasts.

Regarding the remaining parameters, we refer to $K=3$ as a situation of \textit{low complexity} and $K=5$ as \textit{medium complexity}. Furthermore, we consider two interdependence structures: block diagonal and random. The first groups interdependencies in blocks of size $K+1$, while the second allocates them randomly. This, in turn, means that under $K=3$, $N=12$, $M=3$, and \textit{block diagonal}, the complex task---the firm's strategic policy---is \textit{perfectly decomposable}, i.e., there are no cross interdependencies between departments.

Finally, we examine how robust the results are to variations in the turnover rate. We set a baseline scenario of no turnover in which the selection process is only performed once, at $t=1$. We denote this scenario by $\tau=\emptyset$. We compare whether the results for no turnover are robust for a moderate turnover scenario  $(\tau=10)$ and a high turnover scenario $(\tau=1)$.

The remaining parameters are fixed, based on preliminary simulation analyses which showed that altering them did not significantly change the results. These fixed parameters are outlined in Tab. \ref{tab:parameters}. 

For our results, we gather data regarding firm performance, CEO compensation, and the heads of department's compensation---role-wise, rather than for the employees themselves. These measures are normalised to ensure comparability across simulation runs. To answer RQ1, we report firm performance per period, averaged across simulation runs. Regarding RQ2, we gather the cumulative compensation for both the CEO and the heads of department, index both to one at $t=1$, and report their ratio at every period. Thus, this measure captures relative compensation growth, rather than the ratio of absolute pay levels. 

\begin{table}[!htbp]
\centering
\caption{Parameters held fixed across all scenarios.}
\label{tab:parameters}
\begin{tabular}{@{}llll@{}}
\toprule
\textbf{Parameter} & \textbf{Symbol} & \textbf{Value} & \textbf{Submodel} \\
\midrule
Decisions in the complex task      & $N$        & 12   & Task environment \\
Departments                        & $M$        & 3    & Task environment \\
Decisions per department           & $S = N/M$  & 4    & Task environment \\
Employees                          & $A$        & 30   & Head of department selection \\
Periods per run                    & $T$        & 100  & --- \\
Simulation runs                    & $R$        & 1500 & --- \\[3pt]
Estimation noise (s.d., bound)     & $\sigma, \phi$ & 0.1, 0.1 & Head of department selection \\[3pt]
Proposals per department           & \textit{PROP}     & 2    & Decision-making \\
Candidate policies                 & \textit{COMP}     & 2    & Decision-making \\
CEO exploration rate               & $\epsilon$ & 0.25 & Decision-making \\
Benchmark sensitivity              & $\theta$   & 0.9  & Compensation \\
CEO fixed wage                     & $w_{CEO}$  & 0    & Compensation \\
Employee fixed wage                & $w_{emp}$  & 0    & Compensation \\[3pt]
Discovery probability              & $\mathbb{P}$        & 0.25 & Individual learning \\
Forgetting probability             & $\mathbb{Q}$        & 0.25 & Individual learning \\
Memory decay increment             & $\zeta$   & 1    & Individual learning \\
\bottomrule
\end{tabular}
\end{table}
\section{Results}
\subsection{General effects of incentive misalignment on performance}
\begin{figure}[!htb]
 \centering
 \includegraphics[width=\textwidth]{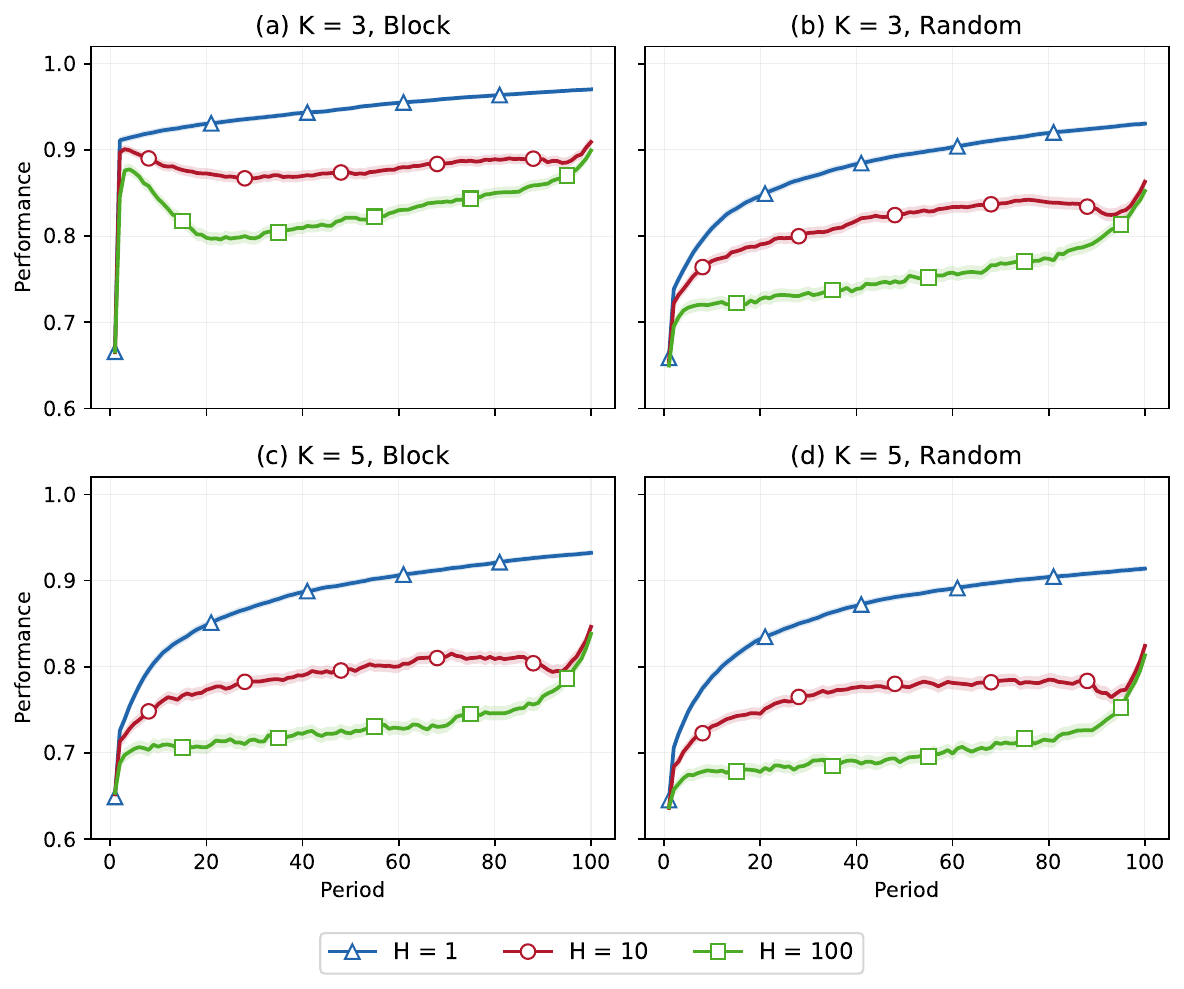}
 \caption{Performance by scenario under no turnover ($\tau=\emptyset$). Shaded bands are 99\% confidence intervals.}
 \label{fig:resultsnoturnover}
\end{figure}

\leavevmode\label{ref:generaleffects}First, we examine the effects of incentive (mis)alignment on performance, without the mediating effect of the head of department turnover rate, i.e., under $\tau=\emptyset$. Figure \ref{fig:resultsnoturnover} shows the four task environments, with each panel reporting the average performance per period over $R=1{,}500$ runs for the three CEO's forecasting horizons considered. Additionally, we report the actual numbers for final performance at $t=100$ and performance averaged over the entire run in Table \ref{tab:noturnover}.

Overall, performance increases under incentive alignment for every task environment studied. Short-term forecasting ($H=1$) reaches a final achieved performance (at $t=100$) between 6.1 and 9.0 percentage points above $H=10$, and between 7.1 and 10.1 points above $H=100$, depending on the task environment considered (see Table \ref{tab:noturnover}). This gap is the smallest for a perfectly decomposable strategic policy ($K=3$ with block interdependencies, panel (a)) and widens as the heads of department's decisions become more interdependent, reaching its maximum for $K=5$ with random interdependencies (panel (d)).

Incentive misalignment does not only reduce the final performance achieved, but also delays the performance improvements that lead to it. Table \ref{tab:noturnover} reports that, averaged over the $t=100$ periods, the performance of $H=1$ exceeds $H=10$ by 6.7 to 9.9 points and $H=100$ by 11.4 to 16.4 points. In both cases, these gaps are larger than the corresponding ones for final performance. Moreover, there are two aspects worth highlighting. First, the two misaligned scenarios end up with similar final performances---0.8 to 1.1 points depending on the scenario---yet they differ by 4.8 to 6.5 points when averaged over the run. Second, because misaligned firms only close part of the gap in the final periods (see Figure \ref{fig:resultsnoturnover}), a firm evaluated before $t=100$ would report a substantially larger performance penalty than the final figures suggest. Thus, we need to look at the complete temporal profile for each scenario in order to understand better the interrelations between incentive misalignment and performance. 

\begin{table}[!htb]
\centering
\caption{Performance under no turnover ($\tau = \emptyset$).}
\label{tab:noturnover}
\begin{tabular}{lcccccc}
\hline
 & \multicolumn{2}{c}{$H = 1$} & \multicolumn{2}{c}{$H = 10$} & \multicolumn{2}{c}{$H = 100$} \\
Task environment & Final & Average & Final & Average & Final & Average \\
\hline
$K=3$, Block & 0.970 & 0.947 & 0.909 & 0.881 & 0.899 & 0.833 \\
$K=3$, Random & 0.931 & 0.882 & 0.863 & 0.815 & 0.852 & 0.754 \\
$K=5$, Block & 0.932 & 0.884 & 0.846 & 0.791 & 0.838 & 0.732 \\
$K=5$, Random & 0.914 & 0.867 & 0.824 & 0.768 & 0.813 & 0.702 \\
\hline
\end{tabular}
\end{table}

The larger gap in mean performance may be explained by the performance trajectories of each forecasting horizon examined. Under incentive alignment, performance improves steadily from the first period, with its rate of improvement declining smoothly over time. By contrast, under $H=10$ and $H=100$, performance improves at a much slower pace, only picking up a faster rate in the last ten periods. During these last ten periods, the CEO's forecasting horizon is truncated by the end of the simulation run---which would be akin to end of tenure or retirement---and its decisions approach those of a myopic CEO ($H=1$).

However, we should note that the perfectly decomposable task environment (panel (a)) slightly departs from these patterns. For $H=1$, performance rises sharply, and then grows at a slow pace for the remaining simulation run. By contrast, for $H=10$ and $H=100$, performance also increases abruptly, then declines over the next 20 to 30 periods, before resuming the slow performance growth and the final surge described above. Incentive misalignment is thus not only costly here but also unstable, as the early gains in performance are partly undone. 

\subsection{Effects under head of department turnover}
\begin{figure}[!htb]
 \centering
 \includegraphics[width=\textwidth]{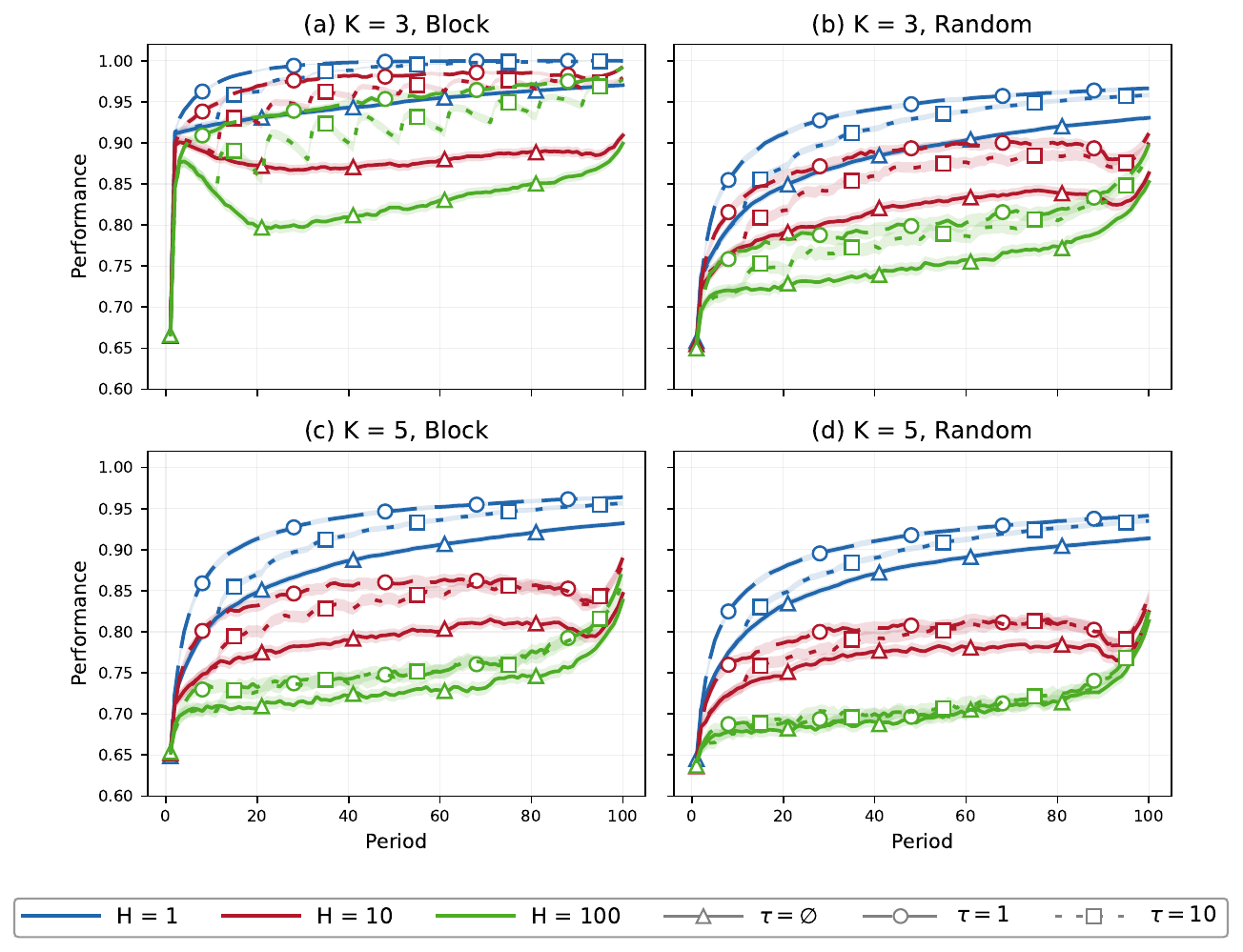}
 \caption{Performance under turnover. Shaded bands are 99\% confidence intervals.}
 \label{fig:resultsturnover}
\end{figure}

\leavevmode\label{ref:turnovereffects}Figure \ref{fig:resultsturnover} shows the results mediated by the rate of head of department turnover $\tau$, while the final and average performances are shown in Table \ref{tab:turnover}.

\begin{table}[!htb]
\centering
\caption{Performance under moderate ($\tau = 10$) and high ($\tau = 1$) turnover.}
\label{tab:turnover}
\begin{tabular}{lcccccc}
\hline
 & \multicolumn{2}{c}{$H = 1$} & \multicolumn{2}{c}{$H = 10$} & \multicolumn{2}{c}{$H = 100$} \\
Task environment & Final & Average & Final & Average & Final & Average \\
\hline
\multicolumn{7}{l}{\textit{Panel A. Moderate turnover ($\tau = 10$)}} \\
$K=3$, Block & 1.000 & 0.982 & 0.979 & 0.955 & 0.977 & 0.919 \\
$K=3$, Random & 0.958 & 0.911 & 0.900 & 0.853 & 0.886 & 0.784 \\
$K=5$, Block & 0.956 & 0.909 & 0.883 & 0.827 & 0.864 & 0.753 \\
$K=5$, Random & 0.935 & 0.885 & 0.845 & 0.786 & 0.826 & 0.711 \\
\hline
\multicolumn{7}{l}{\textit{Panel B. High turnover ($\tau = 1$)}} \\
$K=3$, Block & 1.000 & 0.992 & 0.993 & 0.975 & 0.992 & 0.950 \\
$K=3$, Random & 0.966 & 0.933 & 0.912 & 0.877 & 0.898 & 0.802 \\
$K=5$, Block & 0.964 & 0.932 & 0.890 & 0.845 & 0.877 & 0.757 \\
$K=5$, Random & 0.941 & 0.904 & 0.840 & 0.797 & 0.824 & 0.709 \\
\hline
\end{tabular}
\end{table}

Turnover generally improves both average and final performance. The former increases between 0.8 and 8.6 points for $\tau=10$ and 0.6 to 11.7 points for $\tau=1$, while the latter improvements range between 1.3 and 7.8 points under $\tau=10$ and 1.1 to 9.3 points under $\tau=1$. High turnover weakly dominates moderate turnover, as $\tau=1$ outperforms $\tau=10$ in 10 of the 12 scenarios. In the two exceptions---$K=5$ with random interdependencies, $H=10$ and $H=100$---differences are not significant. Moreover, gains are generally larger for average than for final performance, which suggests that high turnover mainly accelerates the firm's performance improvement, rather than raising the final level attained.

These gains, however, are unevenly distributed across the task environments considered. For a perfectly decomposable strategic policy ($K=3$, block diagonal, panel (a)), the benefits of turnover grow with the CEO's forecasting horizon, from 2.9 points under $H=1$ to 7.8 points under $H=100$ ($\tau=10$). The opposite holds for $K=5$ with random interdependencies (panel (d)), where the gain falls from 2.1 to 1.3 points as the horizon lengthens. The positive effects of replacing the department heads are therefore higher the less interdependent these departments are. Notably, this is the opposite of what the firm requires, as the performance cost of misalignment grows with the interdependencies among departments.

Are the benefits of head of department turnover enough to offset the differences between misaligned and aligned incentives? Results suggest it mostly cannot, with one remarkable exception. Under a perfectly decomposable strategic policy, both $H=10$ and $H=100$ under $\tau=10$ and $\tau=1$ slightly outperform $H=1$ under no turnover $\tau=\emptyset$. These differences grow with more frequent reorganisation, ranging from 0.7 to 0.9 points under $\tau=10$ and reaching 2.2 points under $\tau=1$.

However, this exception disappears when we evaluate the forecasting horizons under the same turnover regime. In all scenarios considered for $\tau=1$ and $\tau=10$, $H=1$ outperforms both $H=10$ and $H=100$ in final and mean performance, signalling both a higher achieved performance and a faster rate of improvement. In case of a perfectly decomposable task, high turnover ($\tau=1$) reduces the performance advantage of $H=1$ from 6.1 to 0.7 points against $H=10$ and from 7.1 to 0.8 against $H=100$. For $K=3$, random interdependencies, and $K=5$, block interdependencies, turnover reduces this gap just around 1 point, while it slightly widens for $K=5$ and random interdependencies. Finally, turnover slightly affects the mean performance differences between $H=10$ and $H=100$, showing a smaller gap for a perfectly decomposable task---3.6 points under $\tau=10$ and 2.4 under $\tau=1$---and a larger gap for $K=5$ and random interdependencies---7.6 points under $\tau=10$ and 8.8 under $\tau=1$. The gaps in final performance, however, remain relatively small in all scenarios considered. 

\subsection{Inequality analysis}
\begin{figure}[!htb]
 \centering
 \includegraphics[width=\textwidth]{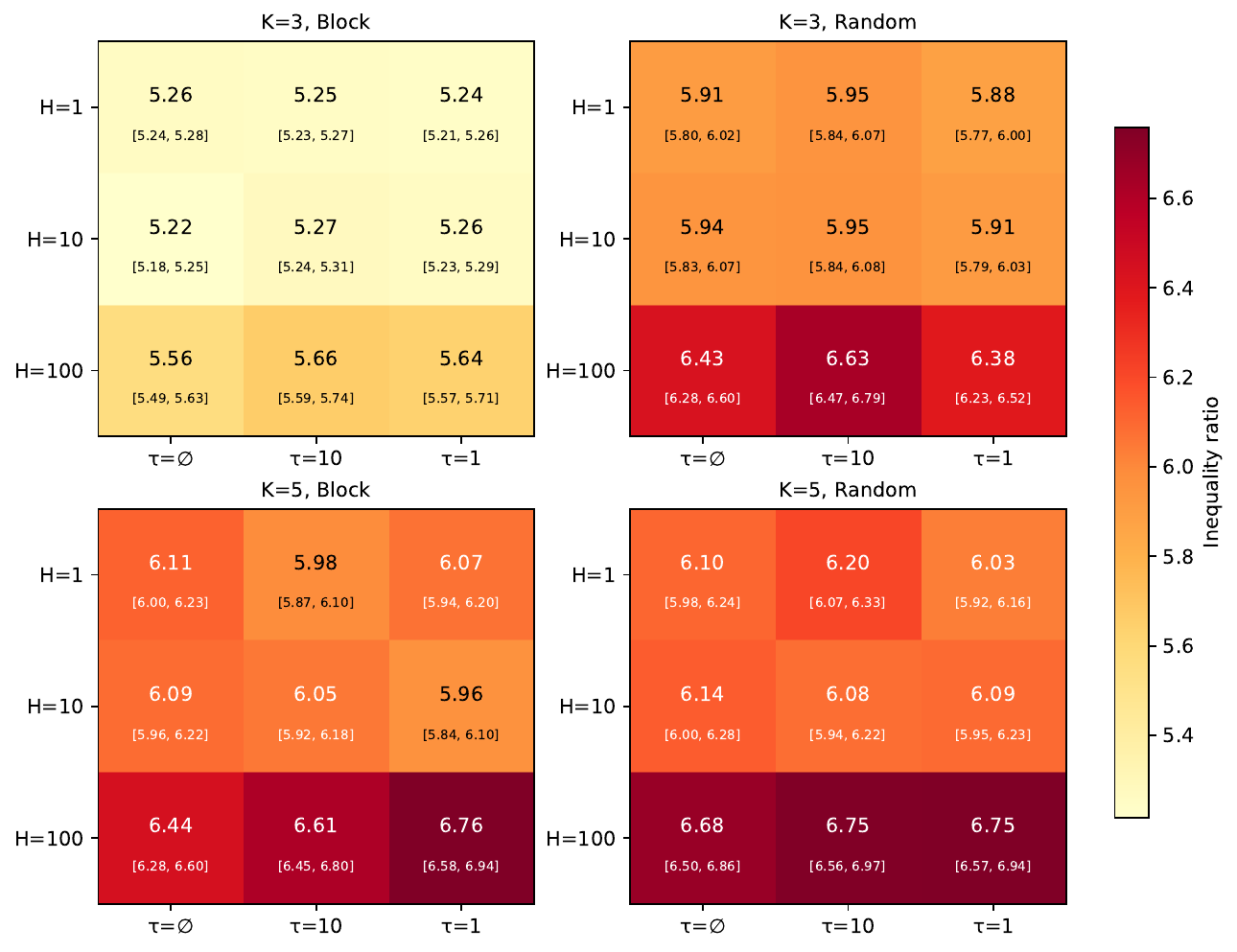}
 \caption{Inequality ratio at $t=100$ by scenario: growth in the CEO's cumulative compensation relative to that of the head of department, both indexed to one at $t=1$. Values in brackets are 95\% bootstrap confidence intervals ($B=1{,}000$).}
 \label{fig:resultsinequality}
\end{figure}

\leavevmode\label{ref:results}Results for the inequality analysis are reported in a heatmap in Figure \ref{fig:resultsinequality}. This figure reports how much faster the CEO's accumulated compensation at $t=100$ has grown compared to the heads of department's, each measured from its own first period $t=1$. For instance, a reported inequality of 5 indicates that the CEO's accumulated pay grew five times faster relative to its starting point than the heads' cumulative compensation. 

Figure \ref{fig:resultsinequality} shows that CEO compensation grows faster under long-term forecasting. For instance, for $H=100$, this ratio ranges between 5.56 and 6.76, depending on the scenario examined. By comparison, $H=1$ yields ratios ranging between 5.24 and 6.20, while $H=10$ exhibits similar numbers (5.22 to 6.14). This inequality is the lowest for a perfectly decomposable strategic policy, and grows as departments become more interdependent. Moreover, turnover appears to have little effect on inequality, with the ratio shifting at most by 0.3 points across scenarios.

To better interpret these results, we compare the ratios obtained with a benchmark inequality ratio. Consider a firm whose performance remains constant over the entire run, and whose CEO always exceeds the performance threshold. In such a firm, the head of department receives the same compensation at every period, so their accumulated compensation grows in proportion to the number of periods elapsed. The CEO's compensation, by contrast, increases at every period due to the weighting factor $1/(T-t+1)$ that favours later-term performance gains. Under constant performance, the inequality ratio would therefore reach approximately 5.18 at $t=100$, which stems solely from the design of the compensation function and is common across all scenarios. Hence, the deviations from this number, as reported in Figure \ref{fig:resultsinequality}, reflect how performance evolves over the run, as performance gains achieved in later periods weigh more heavily in the CEO's compensation than in that of the heads of department. 

\begin{figure}[!htb]
 \centering
 \includegraphics[width=\textwidth]{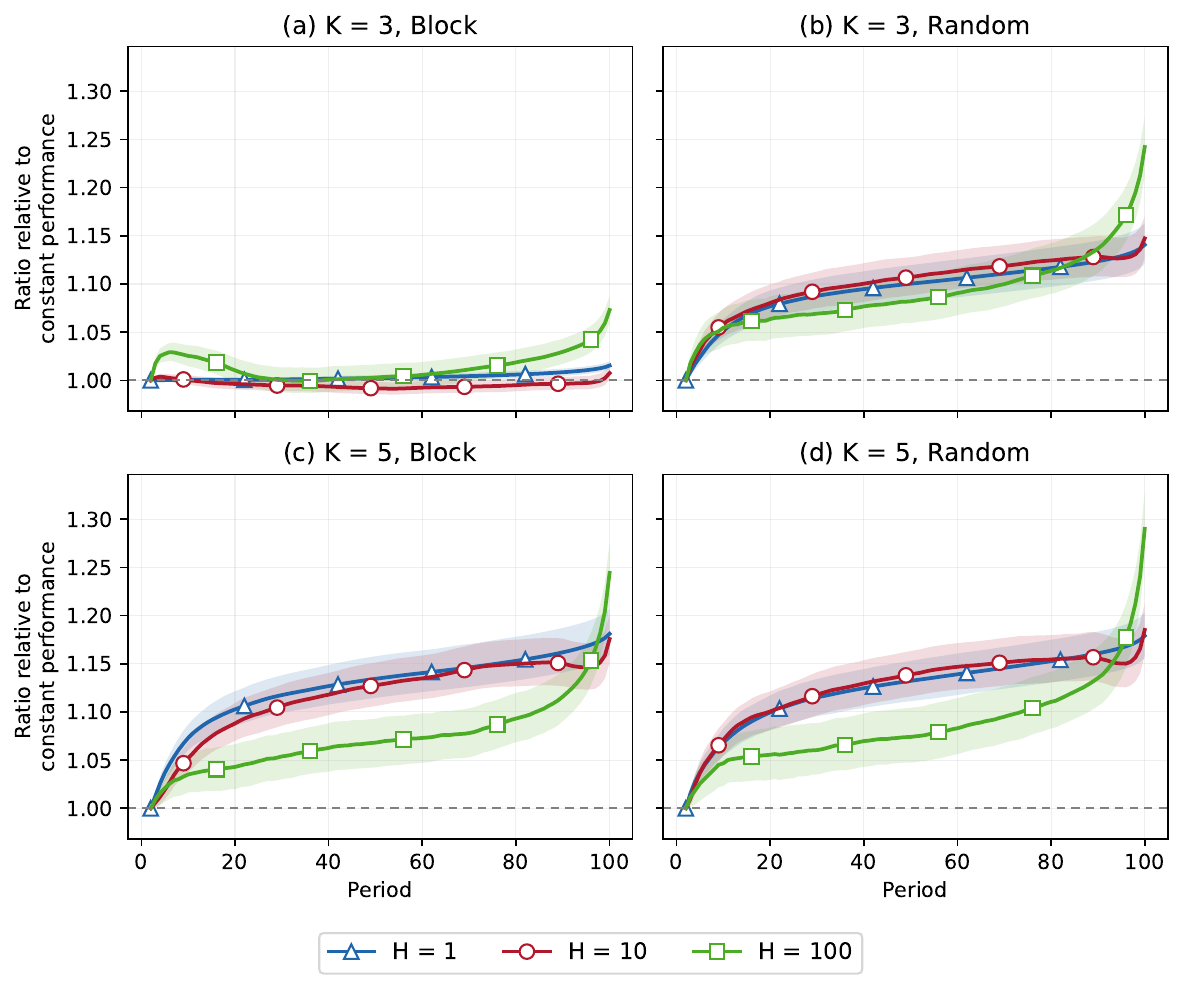}
 \caption{Inequality ratio over time, relative to the ratio that would result from constant firm performance, no turnover $(\tau=\emptyset)$. Shaded bands are 95\% bootstrap confidence intervals ($B=1{,}000$).}
 \label{fig:timeinequality}
\end{figure}

Figure \ref{fig:timeinequality} shows the comparison between the results and the benchmark inequality case over the $T=100$ periods studied and under no turnover $(\tau=\emptyset)$. To isolate the effect of the CEO's and the heads of department's decisions, each ratio is divided by the one that a constant-performance firm would report in the same period. A value of one thus indicates that CEO compensation has grown exactly as fast as its compensation function alone implies, while values above one indicate faster growth driven by performance improvements. 

In all task environments except the perfectly decomposable one (panels (b), (c), and (d)), inequality under $H=1$ and $H=10$ increases gradually as performance improves, and both horizons show a very similar inequality in compensation growth. Additionally, in the perfectly decomposable task (panel (a)), performance reaches a plateau early on, and inequality remains close to the benchmark for $H=1$ and $H=10$. By contrast, $H=100$ exhibits different dynamics. For non-perfectly decomposable tasks (panels (b), (c), and (d)), $H=100$ shows lower inequality for most of the run, reflecting its slower performance growth. However, it eventually overtakes the other horizons in terms of inequality during the last ten periods, which coincides with the late performance recovery that is weighted heavily in the CEO's compensation function. Finally, for a perfectly decomposable task, inequality increases early on, decreasing in the following 20 periods. Then, inequality remains close to the benchmark before rising gradually and then sharply in the final periods. The higher inequality associated with long-term forecasting is therefore produced almost entirely at the end of the run.

Generally, these results suggest that scenarios yielding the lowest average performance---i.e., for $H=100$---are associated with the highest inequality. This mechanism runs through two features of the CEO's compensation function (see Equation \ref{eq:ceo_utility}). First, a firm performing poorly at earlier periods may fail to achieve the performance threshold $\theta\bar{C}^*$ more often, which withholds the variable compensation component. For instance, under $H=100$ the CEO is compensated in 77\% to 96\% of the first ten periods, against 94\% to 100\% under $H=1$. This keeps inequality lower for most of the periods studied. Second, the variable component is weighted such that performance gains at later stages increase compensation more than early improvements. The stagnation followed by late recovery that characterises the $H=100$ performance trajectories translates into a steeper growth in CEO compensation, even though average performance remains below that attained under $H=1$ and $H=10$. This results in CEOs under long-term forecasting earning a comparatively larger share of their cumulative compensation at later periods than CEOs under short- or medium-term forecasting, which sharply increases inequality at later periods and, thus, overall inequality. 

\subsection{Discussion}
\leavevmode\label{ref:discussion}Our findings regarding incentive misalignment are consistent with earlier simulation-based studies on the topic. \cite{Siggelkow2005} report a lower speed of performance growth when departmental and firm-wide incentives diverge, a result shared by \cite{Dosi2003} and \cite{Rivkin2002}. These effects are mediated by the nature of the task environment examined \citep{Wall2016}. Our contribution builds on these results, and is twofold. First, by reporting both final and average performance, we show that incentive misalignment decreases both the speed of performance improvement and its final achieved level. Second, our results suggest that incentive misalignment can emerge in many dimensions---e.g., spatial or temporal---and just one of them is enough to disrupt the firm's search process.

The lower performance generally associated with incentive misalignment may be explained by the inability of the CEO to properly anticipate solution paths that the heads of department will actually learn and propose. This explanation connects our results to what \cite{Aghion1997} describe as "real authority" within organisations: the CEO holds formal authority over the strategic policy implemented, but this implementation is bounded by the heads of department's proposals, which is further bounded by what they learn. The resulting tension is reflected in the CEO's choices. Under misalignment, the CEO adopts policies that may not be the highest-performing in that period, in exchange for projected paths that the organisation may not realise. This is analogous to the mechanism described by \cite{Siggelkow2005}, where a CEO acting under misaligned departmental and firm-wide incentives may reject proposals that would have improved performance, disrupting the heads of department's incremental search. What is novel in our research is that the disruption appears not due to calculation errors, but rather from projecting a performance path that rests on a search process not followed by the CEO's subordinates. 

These results also shed light on the two research questions outlined in the introduction. The first asked how the temporal misalignment of incentives influences the relationship between CEO compensation and firm performance. Our findings suggest that the link between CEO compensation and long-term firm performance appears to be weaker than expected, offering a mechanism which is consistent with prior insights \citep{Bradley2013,Edmans2017,Tosi2004}. We contribute to this literature by showing that incentive misalignment is what weakens the link. Furthermore, our findings dispute the conventional treatment of short-termism in the literature, which is believed to decrease performance \citep{Edmans2017}. By contrast, we find that CEOs who optimise performance step by step, in line with the employees' own horizon, reach higher performance than those pursuing medium- or long-term outcomes. Hence, the results lend support to the scepticism expressed by directors regarding the benefits of making incentives more long-term \citep{Edmans2023}.

However, higher firm performance does not translate into faster growth in CEO compensation. Our results point the other way: CEOs presiding over the worst-performing firms---i.e., for $H=100$---see their accumulated compensation grow the fastest among all scenarios. This decoupling of pay growth from performance is usually read as a signature of rent extraction \citep{Edmans2017,Henriksen2025,Keller2021}. 
Accordingly, the CEOs' pay profile in our model resembles the one produced by ratcheting \citep{Bouwens2011,Bandiera2007}, which can emerge from spreading rewards over the long term to avoid short-termism \citep{Edmans2012,Edmans2016}. Nevertheless, our model reaches these results through a different route. The CEOs in our model hold no power over their own compensation and engage in no extractive behaviour. Additionally, ratcheting is not explicitly included in our model. Instead, our model reproduces similar outcomes to those commonly attributed to strategic manipulation, but which emerge from incentive misalignment alone.

The inequality results also relate to the topic of CEO behaviour at the end of tenure. \cite{Dechow1991} show that CEOs reduce long-term investments, such as in R\&D activities, in their final years in office. \cite{Edmans2012} and \cite{Edmans2016} propose countering this short-termist behaviour by making compensation increasingly sensitive to performance as tenure ends, and contingent on performance after the CEO leaves the firm. The improved performance and increasing inequality present for $H=100$ in our results reflect the first dynamic, but with the opposite consequence. As the end of the run approaches, the forecasting horizon gets narrower, which effectively turns the CEO short-termist, aligns their decisions with the heads of department's incentives, and improves performance. Because CEO compensation is most sensitive to performance in these final periods, this late improvement translates into the sharp surge in inequality observed in Figure \ref{fig:timeinequality}. Hence, in our model, short-termism at the end of tenure improves performance while widening the pay gap, which suggests that extending the CEO's horizon beyond tenure, as \cite{Edmans2012} and \cite{Edmans2016} suggest, may be counterproductive for performance when lower-level decision-makers face short-term incentives.

The double effect identified in the results---increased inequality and poor performance---makes it difficult to justify CEO compensation growth on the grounds of managerial contribution, an argument often discussed in the rent extraction literature \citep{Henriksen2025}. Furthermore, it may also carry costs our model does not capture. Wider inequality, especially under weak performance, may reduce employees' motivation and productivity \citep{Mueller2017}, and directors themselves acknowledge the reputational costs associated with rising CEO pay \citep{Edmans2023}. While these aspects fall outside our scope of analysis, incorporating them would be a natural extension of the CEO-employee perspective employed here.

Finally, we should note that the model reproduces the two patterns described in Section \ref{sec:patterns}. First, incentive alignment is associated with faster performance improvement. Second, CEO compensation grows substantially faster than that of the head of department in every scenario examined.

\subsection{Limitations}
Our results have some limitations worth discussing. First, we assume fixed wages that remain constant across periods and set them equal to zero, so that our analysis concerns the evolution of variable pay, i.e., the component that has grown most over time. Second, and relatedly, our compensation functions are stylised and do not reproduce the full structure of executive and employee pay, which includes stocks, options and vesting schedules. Third, the end of tenure is fixed at $T=100$, which is the end of the simulation run. Since the inequality surge under $H=100$ appears in the final periods, these effects could be read as end-of-tenure effects under a known horizon. Finally, our inequality measure captures CEO compensation growth, rather than differences in levels. A potential extension of this work could involve calibrating the model's parameters with real-world data and relaxing these assumptions, which would allow us to gain further insights into the relationship between CEO compensation design, firm performance and intra-firm inequality.

\section{Conclusion}
This paper pursued a twofold objective. First, we examined how the temporal misalignment between the CEO’s and the employees’ incentives influences the relationship between CEO compensation and firm performance. Second, we traced its implications for intra-firm compensation inequality. To achieve these objectives, we have designed and implemented an ABM based on the NK framework that includes a hierarchical firm, a CEO endowed with imperfect forecasting capabilities, and several heads of department whose compensation rewards only the current performance of their own department. 

Prior research has addressed both questions predominantly from the top down, as the outcome of negotiations between the CEO, the board and the shareholders \citep{Edmans2017}. Our approach, by contrast, dealt with these issues from the bottom up, examining the performance outcomes that emerge from the interactions between the CEO and lower-level employees. This approach to both issues places our study at the intersection of several lines of research, including organisational decision-making, income inequality, and CEO compensation design. Furthermore, it allowed us to depart from the usual shareholder-rent extraction dichotomy. The former explains compensation dynamics through optimal contracting, while the latter does it through managerial power. Our results, by contrast, suggest that neither is required to produce these patterns. Instead, these patterns emerge in our model from the interactions between the CEO’s long-term compensation scheme and the short-term incentives of lower-level decision-makers.

Our results contribute to the literature in three respects. We show that temporal incentive misalignment reduces firm performance, lowering both the level attained and the speed at which the firm improves. By contrast, short-term alignment leads to persistent improvements in performance, even if there is turnover among the firm's decision-makers. Finally, the gap between the growth of CEO compensation and that of the heads of department is widest under long-term forecasting. This association has been traditionally attributed to rent extraction. However, our model reproduces this pattern even in the absence of agents holding power over their own compensation or rent-extracting behaviour.

Taken together, these results identify a tension that executive compensation design cannot avoid. Long-term incentives are meant to align the CEO's decisions with the shareholders' interests. Yet they can misalign the CEO with the heads of department, whose decisions also shape firm performance. Since this misalignment has a double effect, reducing performance and widening inequality, the temporal alignment of incentives with lower-level employees must be considered alongside their level and structure when designing compensation packages.
\begin{credits}
\subsubsection{\discintname}
The author has no competing interests to declare that are relevant to the content of this article.
\end{credits}
\bibliographystyle{jasss}
\bibliography{refs}
\appendix
\setcounter{secnumdepth}{3}
\section{ODD model description}
\subsection{Purpose and patterns}
\leavevmode\label{sec:appendix}The purpose of this model is to examine how the temporal (mis)alignment between the CEO's and the employees' incentive schemes shapes both firm performance and the divergence in compensation growth between the CEO and employees. Our approach departs from the top-down analysis of CEO compensation as an optimal contracting problem between two parties, as in the principal-agent framework \citep{Edmans2017,Hlmstrom1979}. Instead, we conceptualise firm performance and the relative growth in CEO and employee compensation as emergent outcomes of the interactions between multiple boundedly rational decision-makers of a hierarchical organisation.

We assess the validity of this model against two patterns that are often documented in the literature on organisational decision-making and CEO compensation. The first pattern is that incentive alignment between decision-makers is associated with faster organisational performance improvement \citep{Rivkin2002,Siggelkow2005,Wall2016}. The second is the accelerating growth in CEO compensation relative to other employees, including other high-level executives \citep{Edmans2016,Henriksen2025,Mueller2017,Piketty2014}.

\subsection{Entities, state variables, and scales}
\subsubsection{Entities}
\leavevmode\label{sec1}\label{sec:entities}At the individual level, we have two kinds of agents: the "CEO" and the "employees". The CEO is a single agent who sits at the top of the firm's hierarchy. Conversely, the employee population consists of $A$ agents, each pre-assigned to one of the $M$ departments. Taken together, the CEO and the departments form a "firm" (see Sec. \ref{sec:collectives}). The construction of this firm is based on the hierarchical organisational archetype, as outlined in \cite{Siggelkow2005}.

Employees compete with their peers---i.e., with the other employees assigned to each department---to become "head of department" (see Submodel \ref{sec:hod_selection}). The heads of department are the responsible decision-makers for each department. 

Both the CEO and the employees aim to maximise their compensation, although they follow fundamentally different objectives and decision rules (see Sec. \ref{sec:objectives}). To achieve their objectives, the agents of the model need to make decisions, observe the results of their actions, and change (or not) their decisions accordingly. This process implies that both the CEO and the employees interact with a "task environment". In our specific formulation, the task environment is based on the NK performance landscape \citep{Levinthal1997} (see Submodel \ref{sec:nkmodel}). We refer to this complex task as the \textit{firm's strategic policy}, and to any partition of this complex task---i.e., to any subtask---as a \textit{departmental policy}.

\subsubsection{State variables}

{\footnotesize
\setlength{\tabcolsep}{4pt}\renewcommand{\arraystretch}{1.15}
\begin{xltabular}{\textwidth}{@{}N l U D@{}}
\caption{State variables of the model, grouped by entity.}\label{tab:app_state_variables}\\
\toprule
\textbf{Name} & \textbf{Symbol} & \textbf{Units} & \textbf{Description}\\ \midrule
\endfirsthead
\multicolumn{4}{@{}l}{\itshape Table \thetable\ (continued)}\\[2pt] \toprule
\textbf{Name} & \textbf{Symbol} & \textbf{Units} & \textbf{Description}\\ \midrule
\endhead
\midrule \multicolumn{4}{r@{}}{\itshape continued on next page}\\
\endfoot
\bottomrule
\endlastfoot
\multicolumn{4}{@{}l}{\textit{Entity: CEO}}\\
Compensation & $u_t(\mathrm{CEO})$ & Real number. & Compensation received by the CEO at time $t$. Their decision-making process changes according to the compensation received. Changes after the decision-making process.\\
Projected compensation & $\hat{U}_t(\mathbf{d}_t)$ & Real number. & Projected compensation over the $H$-period horizon. Their decision-making process changes according to the compensation received. Changes during the decision-making process.\\
Performance threshold & $\bar{C}^{*}$ & Real number, between 0 and 1. & Performance threshold, randomly drawn from the set of attainable performances at the first period $t=1$. Changes between simulation runs.\\
\midrule
\multicolumn{4}{@{}l}{\textit{Entity: Employees}}\\
Department assignment & $m$ & Integer index, $m \in \{1,...,M\}$. & Employees are randomly assigned to one department at the beginning of each simulation round. Fixed in each simulation round.\\
Role & --- & --- & Employees are selected over time to represent the role of head of department. Changes every $\tau$ periods, during the head of department selection process.\\
Compensation & $u_t(m)$ & Real number. & Compensation received by the employee at time $t$. Their decision-making process changes according to the compensation received. Changes after the decision-making process.\\
Set of known departmental policies & $\Omega_{mt}$ & Set of $S$-dimensional binary vectors. & Set of departmental policies for department $m$ known by the employee. Changes during the individual learning process.\\
Discovery beliefs & $p^{(i)}_{mt}$ & Real number, between 0 and 1. & Belief about discovering solution $i$. Changes during the individual learning process.\\
Forgetting beliefs & $q^{(i)}_{mt}$ & Real number, between 0 and 1. & Belief about forgetting solution $i$. Changes during the individual learning process.\\
\midrule
\multicolumn{4}{@{}l}{\textit{Entity: NK-landscape}}\\
Time & $t$ & Real number, between 0 and $T$. & Time period. At the end of each time period, it is updated to $t=t+1$. Each simulation round ends at $T$.\\
Strategic policy & $\mathbf{d}_t$ & $N$-dimensional vector of binary decisions. & Solution to a complex task formed by $N$ binary decisions, which represents a strategic policy of the firm for a particular period, as decided by the CEO and the heads of department. Changes over time.\\
Decision & $d_i$ & Binary value, 0 and 1. & Individual decision. The strategic policy of the firm is formed by concatenating all binary decisions. Final outcome of the decision-making process.\\
Contribution & $c_i$ & Real number, between 0 and 1. & Each decision has a particular numerical contribution to the overall firm performance. This contribution depends on the value taken by the decision and $K$ other decisions.\\
Firm performance & $C_t(\mathbf{d}_t)$ & Real number, between 0 and 1. & Firm performance, average of all contributions. Each solution $\mathbf{d}_t$ is mapped to its associated performance, forming the performance landscape. Changes after the decision-making process.\\
\end{xltabular}}

\subsubsection{Scales}
\leavevmode\label{sec:scales}The model lacks any spatial dimension. However, we should highlight two structural dimensions that we consider. First, the $N$-dimensional performance landscape has a particular interdependence structure, which reflects the interactions among individual decisions (see Submodel \ref{sec:nkmodel}). Second, the modelled firm has two hierarchical levels: the CEO level and the Departments.

In a temporal sense, we model time as a series of discrete periods denoted by $t$ and ending in $T$. Each period represents a strategic reporting cycle of the firm, such as a quarter or a month. Nonetheless, we deliberately keep the temporal modelling abstract for generalisation purposes. 

\subsection{Process overview and scheduling}
\leavevmode\label{sec:events}The proposed model is structured around four processes. One concerns the model initialisation. Two processes are related to CEO decision-making. In the first, the CEO selects the heads of department. In the second, the CEO---using the heads of department's proposals---implements a particular strategic policy for the firm. Finally, the remaining one concerns the employees' individual learning. Each of these processes corresponds to the sequence of events of the model, as described in Fig.~\ref{fig:events} in Sec. \ref{sec:model}.

The initialisation process takes place before the main sequence of events, and we use it to determine which scenarios to examine (see Sec. \ref{sec:initialisation}). In it, we define the firm environment, the CEO, and the employees according to the state variables described in Tab. \ref{tab:app_state_variables} and the parameters described in Tab. \ref{tab:parameters}. In this initialisation phase, the employees are randomly assigned to one of the $M$ departments. Once this is done, the main sequence of events starts.

The sequence of events starts at $t=1$ and is repeated recursively until it reaches $t=T$. The first event to take place at $t=1$ is the selection of the heads of department. Employees communicate to the CEO their expected contribution to firm performance for the incoming period. The CEO, in turn, names the employees with the highest expected contribution in each department as its head. Thus, during this process, the head of department status is updated synchronously. While this process initiates the loop of events, it is not necessarily repeated at every period $t$. Instead, the frequency of this process is determined by $\tau$. The complete head of department selection process is detailed in Submodel \ref{sec:hod_selection}. 

Next, the decision-making process---i.e., the process of choosing a strategic policy for the firm---occurs. Each head of department sends proposals to the CEO regarding their departmental policy. The CEO gathers and evaluates the information received and implements the strategic policy for the incoming period. Once the decision-making process takes place, the corresponding firm performance is calculated. This also involves the calculation of the compensation for the CEO and the employees (see Sec. \ref{sec:compensation}). Consequently, the firm's strategic policy, firm performance, the CEO's projected and real compensation, and the employees' compensation are synchronously updated. The full decision-making process is outlined in detail in Submodel \ref{sec:decision-making}. 

The employees' individual learning process is the last one to take place at every period. This individual learning process involves three subprocesses that modify the employees' knowledge. First, employees alter the beliefs about what they know according to their achieved compensation. Second, their unused knowledge wears out over time. Third, guided by their beliefs and memory decay, they may discover something new, or forget irrelevant information. Hence, the set of known solutions and the employees' beliefs update individually for each agent, i.e., asynchronously. We detail the individual learning process in Submodel \ref{sec:individual}.

\subsection{Design concepts}
\subsubsection{Basic principles}
\leavevmode\label{sec:design}The model builds on the observed differences in compensation packages between CEOs and lower-rank employees, as outlined in Sec. \ref{sec:stylised}. Specifically, executive compensation in our model is variable, dependent on long-term performance, and downwardly rigid. Conversely, employee compensation depends partly on short-term performance. Details on how these compensation packages are implemented are outlined in Sec. \ref{sec:objectives}.

Another basic principle of this paper can be derived from the dichotomy between CEO and employee compensation: incentive (mis)alignment between decision-makers. Incentive (mis)alignment has long been a key topic of analysis in the literature on CEO compensation and firm performance \citep{Edmans2017}. However, our work moves from the usual shareholder-CEO incentive alignment perspective towards a CEO-employee standpoint (see Sec. \ref{sec:incentive}). 

Incentive (mis)alignment has consequences for the decision-making processes of the CEO and the employees. Additionally, organisational decision-making processes are shaped by rules, hierarchies, and internal bureaucratic processes \citep{Siggelkow2005}. In that sense, we model the internal dynamics of a firm drawing on the hierarchical organisational archetype \citep{Siggelkow2005}. According to this organisational archetype, information flows from lower-level employees to the CEO, who is the final decision-maker of the firm. 

The decision-making environment is based on the NK framework, as adapted by \cite{Levinthal1997} for organisational science. In our setting, a firm faces a complex task---e.g., implementing a specific strategic policy---formed by individual interdependent decisions. Complexity is determined by these interdependencies: as they grow, finding an optimal strategic policy becomes harder. This, in turn, determines one of the key principles of the model: incremental search. The employees and the CEO search the solution space gradually, rather than identifying the optimal strategic policy directly. 

\subsubsection{Emergence}
\leavevmode\label{sec:emergence}The two main outcomes of the model---firm performance and the growth in CEO/employee compensation---emerge from the interactive decision-making processes of the CEO and the employees. Although we define the decision-making rules of each type of agent, these outcomes are mostly determined by the learning process of the employees and the prior history of decision-making in the firm. However, the compensation function of each agent, the NK performance landscape, and the hierarchical structure of the firm are imposed elements.

\subsubsection{Adaptation}
\leavevmode\label{sec:adaptation}The model includes three direct objective-seeking adaptive behaviours, one for each decision point. First, during the heads of department selection process, the employees need to signal to the CEO their expected contribution to firm performance, acting as if they were heads of department (see Submodel \ref{sec:hod_selection}). Second, the heads of department must choose which departmental policies are proposed to the CEO (see Submodel \ref{sec:decision-making}). Third, the CEO needs to implement a strategic policy for the firm.

\subsubsection{Objectives}
\leavevmode\label{sec:objectives}The objective of the employees is to maximise their own compensation. This compensation consists of a fixed wage plus a variable component based on the performance contribution of their department's decisions. However, only employees chosen by the CEO as heads of department receive the variable compensation component. Conversely, regular employees only earn the fixed wage. We define the compensation of the head of department $m$ as:
\begin{equation}\label{eq:agent_utility}
    u_t(m) = w_{emp}+\frac{1}{S} \sum_{n \in \mathbf{d}_m} c_n(\mathbf{d}_t),
\end{equation}
where $w_{emp}$ is the fixed wage and $c_n(\mathbf{d}_t)$ is the performance contribution of decision $d_n$, which depends on the decision itself and $K$ other decisions, and $S=N/M$ is the number of decisions assigned to each department.

The CEO also aims to maximise their compensation, which consists of a fixed wage plus a variable component that depends on long-term performance. This is defined in Equation \ref{eq:appceo_utility}. 

\begin{equation}\label{eq:appceo_utility}
    u_t(CEO) = \begin{cases}
        w_{CEO} \textrm{ if } C(\mathbf{d}_t) \leq \theta\bar{C}^* \\
        w_{CEO} + \frac{C(\mathbf{d}_t)}{T - t+1} \textrm{ if } C(\mathbf{d}_t) > \theta\bar{C}^*. 
    \end{cases}
\end{equation}

The dependence on long-term performance implies that the CEO is incentivised to maximise their long-term compensation, rather than only immediate gains. Hence, we endow the CEO with the capacity to (imperfectly) estimate their compensation over $H$ periods (see Submodel \ref{sec:decision-making}). Consequently, the CEO's objective is to maximise compensation over an $H$-period horizon, and the objective function becomes:

\begin{equation}\label{eq:appceo_objective}
    \hat{U}_t(\mathbf{d}_t) = \sum_{h=0}^{H-1}
    \begin{cases}
        w_{CEO}  \text{ if } C(\mathbf{d}_{t+h}) \leq \theta\bar{C}^{*}, \\[4pt]
        w_{CEO} + \dfrac{C(\mathbf{d}_{t+h})}{T-(t+h)+1} & \text{if } C(\mathbf{d}_{t+h}) > \theta\bar{C}^{*}.
    \end{cases}
\end{equation}

where $w_{CEO}$ is a fixed wage, $\mathbf{d}_{t+h}$ is the policy the CEO projects for period $t+h$, $C(\mathbf{d}_{t+h})$ is firm performance at time $t+h$, and $T - (t+h)+1$ represents the long-term weighting mechanism. Furthermore, $\bar{C}^{*}$ is a performance threshold, while $\theta$ is a sensitivity parameter which scales this performance threshold. These two factors, taken together, determine whether or not the CEO receives the variable compensation component. This is equivalent to a process of  locking the variable compensation component until a threshold performance is reached, which is typical when setting CEO compensation \citep{Edmans2017}.

The difference in the temporal orientation of the employees' and CEO's objective functions is the cornerstone of our model. By modifying the value of parameter $H$, we can alter the alignment of the CEO's and the employees' objectives. For instance, setting $H=1$ effectively means that the CEO is only concerned with immediate compensation maximisation, which aligns their decision-making horizon with the employees' objectives. However, as $H$ increases beyond 1, incentives diverge progressively, since the CEO's forecasting horizon moves even further from the employees' short-term objectives. 

\subsubsection{Learning}
\leavevmode\label{sec:learning}Learning is implemented in our model for all employees, and detailed in Submodel \ref{sec:individual}. It follows a reinforcement learning mechanism with Bayesian updates of beliefs for two processes: discovering and forgetting \citep{Blanco-Fernndez2025}. All agents know at least one departmental strategic policy, i.e., a particular solution to their assigned subtask. At the end of each period, all employees may discover an unknown departmental policy, or forget a known one. These two processes are independent, and occur with the probability $\mathbb{P}$ and $\mathbb{Q}$, respectively. While the probability of either of these events---discovery or forgetting---is fixed, the \textit{likelihood} of forgetting or rediscovering a particular policy changes according to the employees' own beliefs. This means that well-performing policies are less likely to be forgotten and more likely to be discovered, while suboptimal policies are more likely to be forgotten and less likely to be discovered. Heads of department update these beliefs at every period according to the implemented strategic policy and its impact on compensation. When the implemented departmental policy increases (decreases) compensation, a double updating occurs. First, the probability of forgetting this policy decreases (increases). Second, the probability of rediscovering this policy---in case it was previously forgotten---increases (decreases). Finally, we introduce a memory decay factor for all employees, which increases the probability of forgetting a policy the longer it is stored in their memory.

\subsubsection{Prediction}
\leavevmode\label{prediction}Explicit prediction is included in the model. Employees predict their compensation for the incoming period in the head of department selection process, while heads of department do so during the decision-making process (see Submodels \ref{sec:hod_selection} and \ref{sec:decision-making}). These predictions are imperfect, as they include a margin of error (see Eq. \ref{eq:agent_error} in Submodel \ref{sec:hod_selection}). Additionally, the CEO predicts future compensation by simulating candidate strategic policy paths over $H$ periods. This prediction is imperfect, as the CEO cannot foresee the employees' future learning. Hence, a larger horizon $H$ amplifies the potential forecasting errors that may emerge. 

\subsubsection{Sensing}
\leavevmode\label{sec:sensing}The different roles assumed by the agents of the model determine the information received by each of them. The employees' knowledge is limited to the department $m$ they are assigned to. This means that there is no knowledge crossover among departments. Furthermore, they know a limited subset of all potential departmental policies, which is updated over time via learning (see Sec. \ref{sec:learning}). As long as they remain regular employees---and not heads of department---they cannot observe the heads of department's compensation, nor the CEO's. Moreover, they cannot observe each other's known solutions. 

Heads of department are able to see the firm's strategic policy implemented in the previous period---including the policies of the remaining departments---, as they use this information to make their proposed departmental policies. They observe their own compensation, and readjust their beliefs and policies based on this information (see Sec. \ref{sec:learning}). 

The CEO observes the noisy signals of the employees, and uses them to choose the heads of department. Furthermore, the CEO senses all proposals submitted by the heads of department, and builds \textit{COMP} candidate policies from them. Finally, the CEO also observes the immediate performance and compensation associated with each candidate policy, and uses it to imperfectly estimate solution paths over $H$ periods.

\subsubsection{Interaction}
\leavevmode\label{sec:interaction}Interactions are direct between levels and mediated within them. Employees interact with the CEO through their signalling process, whilst the CEO interacts with the employees by choosing the heads of department. Moreover, the heads of department interact with the CEO by sending their proposals. By contrast, the CEO interacts with the heads of department by accepting or rejecting these proposals.

Regarding the within-level interactions, they are mediated by the CEO. Employees interact by competing for the single head of department slot. Additionally, heads of department interact indirectly by affecting each other's contributions---since decisions are interdependent---and, thus, compensation.

\subsubsection{Stochasticity}
\leavevmode\label{sec:stochasticity}Stochasticity is present throughout the model to represent several aspects related to bounded rationality, imperfect information, and variation across simulation runs, all without modelling their root causes. Table \ref{tab:stochasticity} lists every source of randomness, the corresponding random draw, and the aspect of the modelled system it stands for. Each combination of simulation run and scenario is seeded independently, so that landscapes, initial conditions, and behavioural draws vary freely across both.

\begin{xltabular}{\textwidth}{@{}p{2.9cm} p{5.0cm} X@{}}
\caption{Sources of stochasticity, grouped by the stage at which they occur.}
\label{tab:stochasticity}\\
\toprule
\textbf{Element} & \textbf{Random draw} & \textbf{Represents} \\
\midrule
\endfirsthead
\multicolumn{3}{@{}l}{\textit{Table \thetable{} (continued)}}\\
\toprule
\textbf{Element} & \textbf{Random draw} & \textbf{Represents} \\
\midrule
\endhead
\midrule
\multicolumn{3}{r@{}}{\textit{continued on next page}}\\
\endfoot
\bottomrule
\endlastfoot

\multicolumn{3}{@{}l}{\textit{Initialisation (once per simulation run)}}\\[2pt]
Performance contributions
& For each decision $i$ and each configuration of the $K+1$ decisions it depends on, a contribution $c_i$ is drawn independently from $U(0,1)$. Firm performance is $C(\mathbf{d}) = \frac{1}{N}\sum_i c_i$.
& An unknown, rugged task environment whose payoffs cannot be anticipated by the agents. \\[4pt]

Interdependence structure
& In the \textit{random} landscape, the $K$ interdependencies of each decision are drawn at random. In the \textit{block-diagonal} landscape they are fixed by construction and involve no draw.
& Variation in how decision interdependencies are distributed within and across departments. \\[4pt]

Initial strategic policy
& $\mathbf{d}_1$ drawn uniformly from the $2^{N}$ possible solutions.
& The firm's arbitrary starting position on the landscape. \\[4pt]

Performance benchmark
& $\bar{C}^{*}$ drawn uniformly from the $2^{N}$ attainable performance values, then held fixed for the run.
& An externally set performance target, unrelated to the firm's own trajectory, against which CEO compensation is assessed. \\[4pt]

Department assignment
& Each of the $A$ employees is independently assigned to one of the $M$ departments with probability $1/M$. Department sizes are therefore random; assignments leaving a department empty are redrawn.
& The absence of systematic sorting of employees into departments, and the resulting variation in the intensity of competition for the head of department role. \\[4pt]

Initial knowledge
& Each employee is endowed with one departmental policy, drawn uniformly from the $2^{S}$ alternatives.
& Heterogeneous and incomplete prior knowledge among employees. \\[6pt]

\multicolumn{3}{@{}l}{\textit{Head of department selection (every $\tau$ periods)}}\\[2pt]
Signalling noise
& The estimate $u$ that each employee signals is perturbed by $\eta = \mathrm{clip}\!\left(\mathcal{N}(0,\sigma)\cdot u_t(m),\, -\phi,\, \phi\right)$, with $\sigma = \phi = 0.1$.
& Employees' imperfect ability to anticipate their own contribution to firm performance. \\[4pt]

Tie-breaking
& Ties among the highest signallers in a department are resolved uniformly at random.
& The absence of any further criterion for choosing between equivalent candidates. \\[6pt]

\multicolumn{3}{@{}l}{\textit{Decision-making (every period)}}\\[2pt]
Candidate assembly
& Each of the \textit{COMP} candidate policies is built by drawing, for every department, one of the \textit{PROP} forwarded proposals uniformly at random. Candidates may coincide.
& The CEO's limited capacity to evaluate every combination of the proposals received. \\[4pt]

Forward projection
& At each of the $H$ projected steps, the CEO adopts a uniformly random single-decision flip with probability $\epsilon$, instead of the best available single-decision flip.
& The CEO's inability to foresee how the firm will actually move, and hence the imperfection of long-horizon forecasts. \\[4pt]

Tie-breaking
& Among candidates achieving the largest strictly positive improvement over the incumbent policy, one is selected uniformly at random.
& The absence of any further criterion for choosing between equivalent policies. \\[6pt]

\multicolumn{3}{@{}l}{\textit{Individual learning (every period, all employees)}}\\[2pt]
Discovery event
& Occurs with probability $\mathbb{P}$, provided the employee does not already know all $2^{S}$ policies.
& The chance nature of encountering a new departmental policy. \\[4pt]

Policy discovered
& Drawn from the policies the employee does not know, with probability proportional to the normalised discovery belief $p^{(i)}_{mt}$.
& Policies that previously performed well are more likely to be rediscovered than those that did not. \\[4pt]

Forgetting event
& Occurs with probability $\mathbb{Q}$, independently of discovery, provided the employee knows more than one policy.
& The chance nature of losing access to a known policy, with each employee always retaining at least one. \\[4pt]

Policy forgotten
& Drawn from the policies the employee knows, with probability proportional to the normalised forgetting belief $q^{(i)}_{mt}$, which is itself incremented by $\zeta$ in every period the policy is retained.
& Suboptimal policies, and policies held unused for longer, are more likely to be forgotten. \\
\end{xltabular}

\subsubsection{Collectives}
\leavevmode\label{sec:collectives}All collectives in the model are imposed, rather than emergent. Departments are collectives formed by several employees, one of which is selected by the CEO as head of department. They are created by partitioning the complete firm's strategic policy---i.e., the $N$-dimensional complex task faced by the firm---into several departmental policies---i.e., $M$ subtasks. The pool of $A$ employees is then divided by randomly assigning each of them to one of the $M$ departments. Furthermore, departments shape the behaviour of the system in three ways. First, they bound the employees' knowledge. Second, they define the set of peers among whom employees compete for the head of department role. Finally, they represent the unit at which employee compensation is determined (see Sec. \ref{sec:objectives}). 

The $M$ departments, together with the CEO, form a collective firm. This is the collective level at which the strategic policy is implemented and firm performance is realised.

\subsubsection{Observation}
We record firm performance, CEO compensation, and per-department head of department compensation\footnote{The head of department compensation is reported for the role, and not for the individual employee. This means that we observe how much each head of department earns, irrespective of whether they are replaced or not. We take the head of department 1 as a representative for comparison purposes.} for each period and simulation run. These measures are normalised to ensure comparability across simulation runs. We report two sets of results. First, normalised firm performance per period, averaged across all simulation runs. Second, a measure of intrafirm pay inequality: for both the CEO and the head of department, we compute cumulative compensation over time, index each series to one at $t=1$, and report the ratio of the two at every period. This ratio therefore captures the relative growth of CEO and employee compensation over the simulation, rather than the ratio of their pay levels. Together, these results allow us to examine how firm performance and the growth in CEO and employee compensation are jointly determined.

\subsection{Initialisation}
\leavevmode\label{sec:initialisation}The first step at $t=1$ involves setting the main state variables of the model. We generate the \textit{NK} performance landscape (see Submodel \ref{sec:nkmodel}) and draw a random initial $N$-dimensional strategic policy $\mathbf{d}_1$. Afterwards, we uniformly draw a performance benchmark $\bar{C}^{*}$ from the set of attainable performance values. We then assign the $A$ employees to the $M$ departments, and endow each employee with one departmental policy, drawn uniformly from $2^S$ alternatives. Discovery and forgetting beliefs are then uniformly distributed for all alternative departmental policies and all employees. 

Beyond this common initialisation, we vary four parameters to define the scenarios we examine: the interdependencies $K$, the interdependence structure (see Fig. \ref{fig:matrix}), the CEO planning horizon $H$, and the head of department tenure $\tau$. Each scenario is replicated $R=1500$ times. Initialisation occurs at the beginning of each simulation run, so no two runs share the same task environment, starting policy, or allocation of employees to departments.

\subsection{Input data}
\leavevmode\label{sec:input}This model does not use input data to represent time-varying processes. 

\subsection{Submodels}
\leavevmode\label{sec:submodels}The model---including all submodels---is implemented using Python 3.13.5.

\subsubsection{Task environment: NK landscape}
\leavevmode\label{sec:nkmodel}The model builds on the \textit{NK} framework, adapted to reflect the decision-making process at various levels of a hierarchical firm. We have chosen this framework as it has proven to be a valuable and established tool for modelling and examining decision-making in organisations \citep{Levinthal1997,Siggelkow2005,Wall2021,Wall2023}.

The core idea of the \textit{NK} framework is the modelling of organisational decision-making as an incremental search process on a \textit{performance landscape} \citep{Levinthal1997}. In this context, the firm faces a complex task formed by $N$ binary decisions $d_{n} \in \{0,1\}$. This complex task is equivalent, for instance, to the process of finding an optimal strategic policy \citep{Wall2016}. If we follow this example, we can define each possible configuration of the $N$ binary decisions as a particular strategic policy $\mathbf{d}$ of the firm, which has an associated performance $C(\mathbf{d})$. Ultimately, the \textit{performance landscape} is the mapping of each of these solutions to their associated performance.

Organisational research using the \textit{NK} framework typically assumes that agents have cognitive limitations that prevent them from fully processing a complete complex task \citep{Rivkin2002,Siggelkow2005}. As a result, complex tasks are often partitioned into subtasks and assigned to individual decision-makers. This task decomposition reflects organisational decision-making, where tasks are routinely partitioned and allocated to different departments, whose interdependent choices must be coordinated \citep{Siggelkow2005}. In our context, we define $M$ departments, each tasked with finding a particular departmental strategic policy of length $S=N/M$.

Complexity is introduced in the \textit{NK} framework by making the individual decisions \textit{interdependent}. Each decision $d_{n} \in \{0,1\}$ contributes $c_{n}$ to overall performance $C(\mathbf{d})$. However, when decisions are interdependent, contributions do not depend exclusively on their associated decisions. Instead, a contribution to performance depends on its associated decision, plus $K$ other decisions $c_{n}=f(d_{n}, d_{i_1}, \dots, d_{i_K})$,  where $\{i_1, \dots, i_K \} \subseteq \{1, \dots, n-1, n+1, \dots, N \}$ and $0 \leq K \leq N-1$. The higher $K$ is, the more interdependent are the decisions, and the more difficult it becomes to reach an optimal strategic policy.

\begin{figure}[ht]
 \centering
 \includegraphics[width=\textwidth]{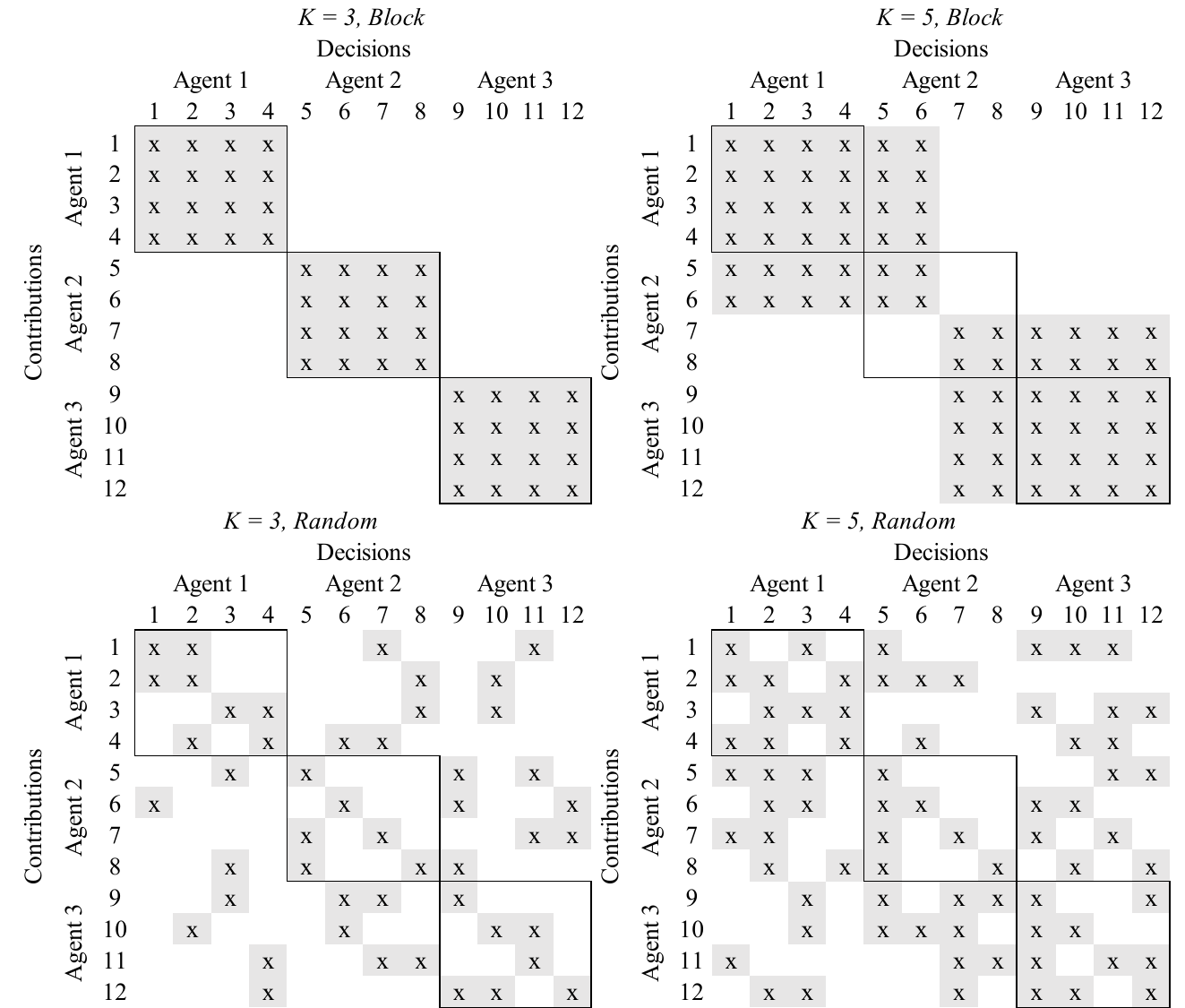}
 \caption{Interdependence matrices. Interdependencies between each contribution and each decision are highlighted in grey with an X.}
 \label{fig:matrix}
\end{figure}

However, the actual \textit{interdependence structure}---i.e., the configuration of which decisions depend on which---is also highly relevant in determining how easy it is to achieve an optimal strategic policy \citep{Rivkin2007}. We consider two structures. In the \textit{block diagonal} structure, interdependencies are grouped into squares of size $K+1$.\footnote{If $S=N/M=K+1$, a block diagonal structure implies that the complex task faced by the firm is \textit{perfectly decomposable}, i.e., there are no cross-interdependencies between departments.} We also consider a \textit{random} structure, in which interdependencies are randomly allocated throughout the task. These two configurations are illustrated in Fig. \ref{fig:matrix}.

\subsubsection{Head of department selection}
\leavevmode\label{sec:hod_selection}To choose the heads of department, the CEO sets up a signalling mechanism at the first period $t=1$, which can be repeated every $\tau$ periods (see Sec. \ref{sec:events}). At time $t$, each employee knows a particular subset of departmental policies ${\mathbf{\Omega}}_{mt}=(\mathbf{d}_{m1},\dots,\mathbf{d}_{mI})$, where $1 \leq I \leq 2^S$ is the number of departmental policies known. For each of these known departmental policies, they estimate their associated compensation. However, two factors prevent employees from perfectly estimating these compensations. First, since they cannot observe other employees' knowledge, they use the strategic policy implemented in the previous period as a placeholder for the remaining departmental policies.\footnote{At $t=1$, a placeholder solution is randomly generated, see Tab. \ref{tab:stochasticity}.}  Second, we endow employees with the capacity to make small errors in their calculations. Thus, following Eq. \ref{eq:agent_utility}, they estimate each contribution according to:

\begin{equation}\label{eq:agent_error}
    \hat{u_t}(\mathbf{d}_m) = w_{emp}+\frac{1}{S} \sum_{n \in \mathbf{d}_m} c_n(\mathbf{d}_t)+\eta ;
\end{equation}

where $\eta$ is an error term  which follows $\eta = \mathrm{clip}\!\left(\mathcal{N}(0,\sigma)\cdot u_t(m),\, -\phi,\, \phi\right)$, with $\sigma = \phi = 0.1$. Once all estimations are done, each employee ranks them in descending order, and signals to the CEO the highest estimated compensation, i.e., $\hat{u_t}\left(\hat{\mathbf{d}}^{(1)}_{mt}\right)$. Then, the CEO selects the highest signaller of each department as its head, breaking any tie randomly. 

\subsubsection{Decision-making}
\leavevmode\label{sec:decision-making}We build the process of choosing a strategic policy for the firm on the hierarchical organisational archetype, as outlined in \cite{Siggelkow2005}. This decision-making process is divided into two steps. In the first step, the heads of department follow the same estimation procedure described in Submodel \ref{sec:hod_selection}, i.e., they estimate all known departmental policies $\mathbf{d}_{mi}\in{\mathbf{\Omega}}_{mt}$ according to Eq. \ref{eq:agent_error}. Once this is done, policies are ranked in descending order according to their estimated compensation. Each head of department then sends to the CEO the top \textit{PROP} departmental policies, i.e., $\hat{\mathbf{d}}^{(1)}_{mt}, \dots , \hat{\mathbf{d}}^{(PROP)}_{mt} \in \Omega_{mt}$.

Once the proposals are gathered, the CEO constructs \textit{COMP} candidate strategic policies by randomly concatenating each head of department's proposals, such that $\mathbf{d}^{(j)}_t := \hat{\mathbf{d}}^{(k^{(j)}_1)}_{1t} \frown \dots \frown \hat{\mathbf{d}}^{(k^{(j)}_M)}_{Mt}$, where the proposals are chosen with a uniform distribution $k^{(j)}_m \sim\mathcal{U}\{1,\dots,PROP\}$. Then, the CEO assesses the \textit{COMP} candidates according to their projected compensation over the $H$ periods, as detailed in Sec. \ref{sec:objectives}, Eq. \ref{eq:ceo_objective}. This forecasting process works as follows: During the $H$ following periods, the CEO observes which alternatives to the current policy exist by flipping the value of one decision, and chooses the one that improves compensation the most. However, at each of the $H$ projected steps, the CEO can also adopt a uniformly random single-decision flip with probability $\epsilon$, instead of the best available single-decision flip. After the \textit{COMP} alternatives have been examined, the CEO chooses the alternative that yields the highest projected compensation over the $H$-period horizon $\mathbf{d}^*_{t}$, and evaluates it against the current implemented policy $\mathbf{d}_{t-1}$---also in the incoming $H$ periods. If the candidate policy improves compensation over the $H$-period horizon, it is implemented $\mathbf{d}_{t}=\mathbf{d}^*_{t}$, while the status quo is maintained if compensation is not projected to improve $\mathbf{d}_{t}=\mathbf{d}_{t-1}$. 

\subsubsection{Compensation}
\leavevmode\label{sec:compensation}Once the strategic policy is implemented, the agents receive their corresponding compensation. Regular employees receive a fixed wage $w_{emp}$. Heads of department are additionally rewarded by receiving the variable component of their compensation $\frac{1}{S} \sum_{n \in \mathbf{d}_m} c_n(\mathbf{d}_t)$, as outlined in Sec. \ref{sec:objectives}, Eq. \ref{eq:agent_utility}. 

The CEO also receives compensation immediately after the decision-making process. The CEO's compensation function is based on the optimal contracting function proposed by \cite{Edmans2016} and several principles of executive pay established in Sec. \ref{sec:stylised}, namely long-term performance dependence and downward rigidity. It consists of a fixed wage, plus a variable component that rewards performance achieved in later periods, and is defined in Equation \ref{eq:appceo_utility}. The preference for long-term performance is represented in the denominator $T - t+1$, which is effectively a weighting factor. Moreover, downward rigidity is operationalised by setting the floor at $w_{CEO}$, which ensures that the CEO is not punished for performance decreases.

\subsubsection{Individual learning}
\leavevmode\label{sec:individual}The individual learning submodel of each employee is divided into three subsubmodels, each representing one of the three aspects of individual learning outlined in Sec. \ref{sec:learning}. 

\textbf{Discovering and forgetting.} By discovering and forgetting departmental policies over time, employees alter their subset ${\Omega}_{mt}$. Discovery refers to the process of finding a currently-unknown departmental policy. This process occurs at the end of each period $t$---see Fig. \ref{fig:events}---with probability $\mathbb{P}$. In addition, employees may also forget a solution currently known, which also occurs at the end of each period with probability $\mathbb{Q}$. We set $\mathbb{P}=\mathbb{Q}$, so the learning process is unbiased toward discovering or forgetting. 

\textbf{Belief update.} Employees may discover or forget policies at a fixed rate, but the probability distribution with which the solutions are discovered or forgotten is not fixed. Instead, it is based on a \textit{reinforcement learning} process by which the heads of department observe the results of their actions and adjust their expectations accordingly. For an employee of department $m$, the belief for discovering solution $\mathbf{d}^m_{i}$, $p^{(i)}_{mt}$, is defined as

\begin{equation}\label{eq:learn}
    p^{(i)}_{mt} = \frac{\alpha^{(i)}_{mt}}{\alpha^{(i)}_{mt}+\beta^{(i)}_{mt}}.
\end{equation}

At $t=1$, all departmental policies have equal beliefs, i.e., departmental policies initially follow a uniform probability distribution.\footnote{In other words, if discovery is activated at $t$, all unknown policies have the same chance of being discovered.}. This means that we initially set $\alpha^{(i)}_{m1}=\beta^{(i)}_{m1} = 1$, so that $p^{(i)}_{m1}=0.5$. After implementing a certain departmental policy $\mathbf{d}^m_{i}$, the head of department $m$ receives a certain compensation, following Eq. \ref{eq:agent_utility}. Depending on the outcome, this head of department $m$ may update their beliefs in one way or another regarding the departmental policy that was just applied.\footnote{Since the remaining employees' compensation is fixed, the implemented strategic policy is irrelevant to their gains, so they cannot update their beliefs accordingly.} This updating process follows a Bayesian updating rule. If compensation increases, $\alpha^{(i)}_{m}$ increases by one. This increases the probability of discovering solution $\mathbf{d}^m_{i}$. By contrast, if policy $\mathbf{d}^m_{i}$ decreases compensation, $\beta^{(i)}_{m}$ will increase by one. The probability of discovering solution $\mathbf{d}^m_{i}$ then decreases as a result. Thus, the updating process for discovering solutions follows Eqs. \ref{eq:learn_a} and \ref{eq:learn_b}.

\begin{subequations}\label{eq:updatelearn}
    \begin{equation}\label{eq:learn_a}
        \alpha^{(i)}_{mt} =
            \begin{cases}
                \alpha^{(i)}_{m\{t-1\}}+1 \textrm{ if } u_t(m) \geq u_{t-1}(m),\\
                \alpha^{(i)}_{m\{t-1\}} \textrm{ otherwise.}
            \end{cases}
    \end{equation}
    \begin{equation}\label{eq:learn_b}
        \beta^{(i)}_{mt} =
            \begin{cases}
                \beta^{(i)}_{m\{t-1\}}+1 \textrm{ if } u_{t}(m) < u_{t-1}(m),\\
                \beta^{(i)}_{m\{t-1\}} \textrm{ otherwise.}
            \end{cases}
    \end{equation}
\end{subequations}

Additionally, employees also hold beliefs for forgetting solutions. For an employee of department $m$, the belief for forgetting solution $\mathbf{d}^m_{i}$, $q^{(i)}_{mt}$, is defined as

\begin{equation}\label{eq:forget}
    q^{(i)}_{mt} = \frac{\lambda^{(i)}_{mt}}{\lambda^{(i)}_{mt}+\delta^{(i)}_{mt}}~.
\end{equation}

As with the case of discovery, we set equal beliefs for all known solutions at $t=1$, such that $\lambda^{(i)}_{m1}=\delta^{(i)}_{m1} = 1$ and $q^{(i)}_{m1}=0.5$. The updating process for these beliefs also follows a Bayesian updating rule. In this case, each time a departmental policy $\mathbf{d}^m_{i}$ is implemented and compensation increases as a result, $\delta^{(i)}_{m}$ increases by one, and the probability of forgetting the policy $\mathbf{d}^m_{i}$ decreases. However, $\lambda^{(i)}_{m}$ increases by one if policy $\mathbf{d}^m_{i}$ reduces the head of department $m$'s compensation, which, in turn, increases the probability of forgetting this policy.

\begin{subequations}\label{eq:updateforget}
    \begin{equation}\label{eq:forget_a}
        \lambda^{(i)}_{mt}=
            \begin{cases}
                \lambda^{(i)}_{m\{t-1\}}+1 \textrm{ if } u_{t}(m) < u_{t-1}(m),\\
                \lambda^{(i)}_{m\{t-1\}} \textrm{ otherwise.}
            \end{cases}
    \end{equation}
    \begin{equation}\label{eq:forget_b}
        \delta^{(i)}_{mt}=
            \begin{cases}
                \delta^{(i)}_{m\{t-1\}}+1 \textrm{ if } u_{t}(m) > u_{t-1}(m),\\
                \delta^{(i)}_{m\{t-1\}} \textrm{ otherwise.}
            \end{cases}
    \end{equation}
\end{subequations}

\textbf{Memory decay.} Additionally, we endow all employees---heads of department or not---with a memory decay factor $\zeta$. This reflects that employees tend to forget policies that have been stored in their memory $\Omega_{mt}$ for a long time, especially if they have not been implemented successfully. This factor alters the employees' forgetting beliefs, increasing the probability of forgetting a particular departmental policy over time. In mathematical terms, we define this memory decay factor as follows: $\forall \mathbf{d}^m_{i} \in \Omega_{mt}: \lambda^{(i)}_{mt} = \lambda^{(i)}_{m\{t-1\}}+\zeta$. 

\section{Code availability}
The code for the model is available at CoMSES Net: \url{https://www.comses.net/codebases/21a00322-604b-4a59-88a5-f978a1f4be91/releases/1.0.0/}.

\end{document}